\documentclass[aps,preprint,showpacs,nofootinbib,showkeys,tightenlines]{revtex4}

\usepackage{graphicx}  %
\usepackage{amsmath}
\usepackage{bm}  %

\newcommand{\bea}{\begin{eqnarray}}
\newcommand{\eea}{\end{eqnarray}}
\newcommand{\beq}{\begin{equation}}
\newcommand{\eeq}{\end{equation}}
\newcommand{\bqa}{\begin{eqnarray}}
\newcommand{\eqa}{\end{eqnarray}}

\def\mqo2{{\!\!\!}}

\usepackage{relsize}
\def\babar{\mbox{\slshape B\kern-0.1em{\smaller A}\kern-0.1em
    B\kern-0.1em{\smaller A\kern-0.2em R}}}

\begin{document}

\title{
Born-Oppenheimer  Approximation 
for the $\bm{XYZ}$ Mesons}
\author{Eric Braaten}
\affiliation{Department of Physics,
         The Ohio State University, Columbus, OH\ 43210, USA\\}
\author{Christian Langmack}
\affiliation{Department of Physics,
         The Ohio State University, Columbus, OH\ 43210, USA\\}
\author{D.~Hudson Smith}
\affiliation{Department of Physics,
         The Ohio State University, Columbus, OH\ 43210, USA\\}
\date{\today}

\begin{abstract}
Many of the $XYZ$ mesons discovered in the last decade can be 
identified as bound states in Born-Oppenheimer (B-O) potentials 
for a heavy quark and antiquark. 
They include quarkonium hybrids, which are bound states in 
excited flavor-singlet B-O potentials, and quarkonium tetraquarks, 
which are bound states in flavor-nonsinglet B-O potentials. 
We present simple parameterizations of the deepest flavor-singlet B-O potentials.
We infer the deepest  flavor-nonsinglet B-O potentials from
lattice QCD calculations of static adjoint mesons.
Selection rules for hadronic transitions are used to identify $XYZ$ mesons
that are candidates for ground-state energy levels in the B-O potentials
for charmonium hybrids and tetraquarks.
The energies of the lowest-energy charmonium hybrids are predicted 
by using the results of lattice QCD calculations to calculate the energy splittings 
between the ground states of different B-O potentials and using the  
Schroedinger equation to determine the splittings 
between energy levels within a B-O potential.
\end{abstract}

\smallskip
\maketitle

\section{Introduction}
\label{sec:intro}
The $XYZ$ mesons are unexpected mesons 
discovered during the last decade that contain 
a heavy quark-antiquark pair and are above the open-heavy-flavor threshold.  
Some of the more surprising of these $XYZ$ mesons are
\begin{itemize}
\item
$X(3872)$, 
discovered by the Belle Collaboration in 2003 \cite{Choi:2003ue}.  
It has comparable branching fractions into $J/\psi\, \rho$ 
and $J/\psi\, \omega$, implying a severe violation of isospin symmetry.
\item
$Y(4260)$, 
discovered by the BaBar Collaboration in 2005 \cite{Aubert:2005rm}.  
It has $J^{PC}$ quantum numbers $1^{--}$, but it is produced very weakly 
in $e^+e^-$ annihilation.
\item
$Y(4140)$, 
discovered by the CDF Collaboration in 2009 \cite{Aaltonen:2009tz}.  
It decays into $J/\psi\, \phi$, which suggests that it might be a tetraquark 
meson with constituents $c \bar c s \bar s$.
\item
$Z_b^+(10610)$ and $Z_b^+(10650)$, 
discovered by the Belle Collaboration in 2011 \cite{Belle:2011aa}.  
They both decay into $\Upsilon\, \pi^+$, 
which implies that they must be tetraquark mesons with constituents 
$b\bar b u \bar d$.
\item
$Z_c^+(3900)$, 
discovered by the BESIII Collaboration in 2013 \cite{Ablikim:2013mio}.  
It decays into $J/\psi\, \pi^+$, which implies that it must be a tetraquark meson 
with constituents $c\bar c u \bar d$.
\end{itemize}
An updated list of the $XYZ$ mesons as of August 2013 was given in 
Ref.~\cite{Bodwin:2013nua}.
In the  $c \bar c$ meson sector, the list consisted of
15 neutral and 4 charged states.
In the  $b \bar b$ meson sector, the list consisted of
1 neutral and 2 charged states.

More than a decade has elapsed since the discovery of the $X(3872)$,
and no compelling explanation for the pattern of $XYZ$ mesons
has emerged.
Simple constituent models for the $XYZ$ mesons can be 
classified according to their constituents and how they are clustered
within the meson.  Those that have been proposed include
\begin{itemize}
\item
{\bf conventional quarkonium}, which
consists of a color-singlet heavy quark-antiquark pair:  $(Q \bar Q)_1$,
\item
{\bf quarkonium hybrid meson}, which
consists of a color-octet $Q \bar Q$ pair
to which a gluonic excitation is bound:  $(Q \bar Q)_8 + g$,
\item
{\bf compact tetraquark} \cite{Vijande:2007rf}, which
consists of a $Q \bar Q$ pair and a light quark $q$ and  antiquark $\bar q$
bound by inter-quark potentials 
into a color singlet:  $(Q \bar Q q \bar q)_1$,
\item
{\bf meson molecule} \cite{Tornqvist:1993ng}, which
consists of color-singlet $Q \bar q$  and $\bar Q q$  mesons
bound by hadronic interactions: $(Q \bar q)_1 + (\bar Q q)_1$,
\item
{\bf diquark-onium} \cite{Drenska:2010kg}, which
consists of a color-antitriplet $Q q$  diquark 
and a color-triplet $\bar Q \bar q$  diquark
bound by the QCD color force:  $(Q q)_{\bar 3} + (\bar Q \bar q)_3$, 
\item
{\bf hadro-quarkonium} \cite{Dubynskiy:2008mq}, which
consists of a color-singlet $Q \bar Q$  pair
to which a color-singlet light-quark  pair is bound by residual QCD forces:
$(Q \bar Q)_1 + (q \bar q)_1$.
An essentially equivalent model is a quarkonium
and a light meson bound by hadronic interactions.
\item
{\bf quarkonium adjoint meson}  \cite{Braaten:2013boa}, which
consists of a color-octet $Q \bar Q$  pair
to which a light quark-antiquark pair is bound:
$(Q \bar Q)_8 +  (q \bar q)_8$.
\end{itemize}
All of these are possible models for neutral $XYZ$ mesons.
The last five are possible models for charged $XYZ$ mesons.
None of these models has proven to be very predictive for the  
pattern of  $XYZ$ mesons.  They are all essentially phenomenological
models whose only connection with the fundamental field theory QCD
is that they use degrees of freedom from QCD.
It would be desirable to have a single theoretical framework 
based firmly on QCD that describes all the $XYZ$ mesons.  One possibility 
for such a framework is the Born-Oppenheimer (B-O) approximation. 

The B-O approximation is used in atomic 
and molecular physics to understand the binding 
of atoms into molecules \cite{Born-Oppenheimer}. 
It exploits the large ratio of the time scale for the motion 
of the atomic nuclei to that for the electrons,
which is a consequence of the large ratio of the mass 
of a nucleus to that of the electron. 
The electrons respond almost instantaneously to the motion of the nuclei. 
Their instantaneous configuration is determined by the positions of the nuclei, 
which can be approximated by static sources for the electric field. 
The energy of the electrons combined with the repulsive 
Coulomb energy of the nuclei defines a Born-Oppenheimer (B-O) potential. 
The B-O approximation to the energy levels 
of the molecule are the energy eigenvalues of 
the Schroedinger equation in that potential.

The B-O approximation for $Q \bar Q$ mesons in QCD
was developed by Juge, Kuti, and Morningstar \cite{Juge:1999ie}. 
It exploits the large ratio of the time scale for the motion 
of the $Q$ and $\bar Q$ to that for the evolution of gluon fields,
which is a consequence of the large ratio 
of the heavy-quark mass $m_Q$ to the nonperturbative momentum scale 
$\Lambda_{\rm QCD}$ associated with the gluon field. 
The gluon field responds 
almost instantaneously to the motion of the $Q$ and $\bar Q$. 
Its instantaneous configuration 
is determined by the positions of the $Q$ and $\bar Q$, 
which can be approximated by static color sources. 
The energy of the gluon field defines a 
B-O potential $V_\Gamma(r)$ 
that depends on the separation $r$ of the $Q$ and $\bar Q$
and on the quantum numbers $\Gamma$ for the gluon field in the presence of 
static $Q$ and $\bar Q$ sources. 
The motion of the $Q$ and $\bar Q$ can be described by the 
Schroedinger equation with potential $V_\Gamma(r)$. 
In the B-O approximation, $Q \bar Q$ mesons 
are energy levels of 
the Schroedinger equation in the B-O potentials.
The energy levels in the ground-state potential are conventional quarkonia.  
The energy levels in the excited-state potentials are quarkonium hybrids.  

Juge, Kuti, and Morningstar calculated many of the B-O potentials 
using quenched lattice QCD,
in which light-quark loops are omitted \cite{Juge:1999ie}.  
They calculated the spectrum of bottomonium hybrids
by solving the Schroedinger equation in the B-O potentials.
They also calculated some of the bottomonium hybrid energies 
using lattice NRQCD.  The quantitative agreement between
the predictions of the B-O approximation
and lattice NRQCD provided convincing evidence for the existence of
quarkonium hybrids in the hadron spectrum of QCD.

For QCD with light quarks, 
the B-O potentials can be defined as the energies of 
stationary configurations of the gluon and light-quark fields whose flavor quantum numbers 
are singlet in the presence of static $Q$ and $\bar Q$ sources.
In Ref.~\cite{Braaten:2013boa}, it was pointed out that B-O potentials
can also be defined by the energies of stationary configurations
of  light-quark and gluon fields that have flavor-nonsinglet quantum numbers.
The energy levels of a $Q \bar Q$ pair in such a flavor-nonsinglet potential
are quarkonium tetraquarks.
Several of the simple constituent models for quarkonium tetraquarks
itemized above can be identified with specific regions of the Born-Oppenheimer 
wavefunction for the $Q \bar Q$ pair.
When the separation of the $Q \bar Q$ pair
is much smaller than the spatial extent of the light-quark and gluon fields,
the system resembles a quarkonium adjoint meson $(Q \bar Q)_8 + (q \bar q)_8$.
When the $Q$ and $\bar Q$ are well separated, the system resembles
a meson molecule $(Q \bar q)_1 + (\bar Q q)_1$ if the light quark is localized
near the $\bar Q$
and it resembles diquark-onium $(Q q)_{\bar 3} + (\bar Q \bar q)_3$  
if the light quark is localized near the $Q$.

In this paper, we apply the B-O approximation for quarkonium 
hybrids and tetraquarks to the $XYZ$ mesons.
In Section~\ref{sec:XYZ}, we list the $XYZ$ mesons that have been observed so far.
In Section~\ref{sec:BOpotential}, we discuss the B-O potentials 
for quarkonium hybrids and tetraquarks.
We present accurate parametrizations of the deepest hybrid B-O potentials,
and we infer the deepest tetraquark B-O potentials from lattice QCD calculations
of static adjoint mesons.
In Section~\ref{sec:SchrEq}, we apply the B-O approximation 
to quarkonium hybrid and tetraquark mesons.
We derive selection rules for hadronic transitions between Born-Oppenheimer 
configurations and use them to identify $XYZ$ mesons that are candidates 
for ground-state energy levels of charmonium hybrids and tetraquarks.
In Section~\ref{sec:lattice}, we describe lattice QCD calculations of 
$c \bar c$ and $b \bar b$ mesons and discuss their implications for 
the B-O approximation.
In Section~\ref{sec:pheno}, we predict the lowest energy levels of 
charmonium hybrids by combining results from lattice QCD with 
energy splittings from solutions of the Schoroedinger equation in B-O potentials.
The outlook for developing the B-O approximation
into a systematic theory of the $XYZ$ mesons is discussed in Section~\ref{sec:outlook}.

\section{$\bm{XYZ}$ Mesons}
\label{sec:XYZ}

\begin{table}[tb]
\begin{center}
\begin{tabular}{cllcll}
 State & $M$~(MeV) & $\Gamma$~(MeV) & ~$J^{PC}$~ & Decay modes & 
     1$^{\rm st}$~observation \\
\hline
~ $X(3823)$~ & 3823.1$\pm$1.9 & $<24$ &
    $?^{?-}$
    & $\chi_{c1}\gamma$ &
    Belle~2013 \\
$X(3872)$& 3871.68$\pm$0.17~ & $<1.2$ &
    $1^{++}$
    & $J/\psi\, \pi^+\pi^-$, $J/\psi\, \pi^+\pi^-\pi^0$~ &
    Belle~2003 \\
& & & & $D^0 \bar D^0 \pi^0$, $D^0 \bar D^0 \gamma$ &  \\
& & & & $J/\psi\, \gamma$, $\psi(2S)\, \gamma$ &  \\
$X(3915)$ & $3917.5\pm1.9$ & 20$\pm 5$ & $0^{++}$ &
    $J/\psi\, \omega$, ($\gamma \gamma$) &
    Belle~2004 \\ 
$\chi_{c2}(2P)$ & $3927.2\pm2.6$ & 24$\pm$6 & $2^{++}$ &
     $D\bar{D}$, ($\gamma \gamma$) &
     Belle~2005 \\ 
$X(3940)$ & $3942^{+9}_{-8}$ & $37^{+27}_{-17}$ & $?^{?+}$ &
     $D^* \bar D$, $D \bar D^*$ &
     Belle~2007 \\
$G(3900)$ & $3943\pm21$ & 52$\pm$11 & $1^{--}$ &
     $D \bar D$, ($e^+ e^-$)&
     \babar~2007 \\ 
     $Y(4008)$ & $4008^{+121}_{-\ 49}$ & 226$\pm$97 & $1^{--}$ &
     $J/\psi\, \pi^+\pi^-$, ($e^+ e^-$) &
     Belle~2007 \\
$Y(4140)$ & $4144.5\pm2.6$  & $15^{+11}_{-\ 7}$ & $?^{?+}$ &
     $J/\psi\, \phi$ &
     CDF~2009 \\
$X(4160)$ & $4156^{+29}_{-25} $ & $139^{+113}_{-65}$ & $?^{?+}$ &
     $D^* \bar D^*$ &
     Belle~2007 \\
$Y(4220)~$ & $4216 \pm 7$ & $39 \pm 17$ & $1^{--}$ &
     $h_c(1P)\, \pi^+\pi^-$,  ($e^+ e^-$)&
    BESIII~2013 \\ 
$Y(4260)$ & $4263^{+8}_{-9}$ & 95$\pm$14 & $1^{--}$ &
     $J/\psi\, \pi^+\pi^-$, $J/\psi\, \pi^0\pi^0$ &
     \babar~2005 \\ 
& & & & $Z_c(3900)\, \pi$, ($e^+ e^-$) &  \\
$Y(4274)$ & $4274.4^{+8.4}_{-6.7}$ & $32^{+22}_{-15}$ & $?^{?+}$ &
     $J/\psi\, \phi$ &
     CDF~2010 \\
$X(4350)$ & $4350.6^{+4.6}_{-5.1}$ & $13.3^{+18.4}_{-10.0}$ & ~0/2$^{++}$~ &
     $J/\psi\, \phi$, ($\gamma \gamma$) &
     Belle~2009 \\ 
$Y(4360)$ & $4361\pm13$ & 74$\pm$18 & $1^{--}$ &
     $\psi(2S)\, \pi^+\pi^-$, ($e^+ e^-$) &
     \babar~2007 \\ 
     $X(4630)$ & $4634^{+\ 9}_{-11}$ & $92^{+41}_{-32}$ & $1^{--}$ &
     $\Lambda_c^+ \Lambda_c^-$, ($e^+ e^-$) &
     Belle~2007 \\
$Y(4660)$ & 4664$\pm$12 & 48$\pm$15 & $1^{--}$ &
     $\psi(2S)\, \pi^+\pi^-$, ($e^+ e^-$) &
     Belle~2007 \\ 
\hline
\end{tabular}
\end{center}
\caption{ Neutral $c\bar{c}$ mesons above the $D \bar{D}$ threshold
discovered since 2003.
Neutral isospin partners of charged $c \bar  c$ mesons are not listed.
The ``decay modes'' in parenthesis, ($e^+ e^-$) and ($\gamma \gamma$),
are actually production channels.} 
\label{tab:ccneutral}
\end{table}

Lists of the $XYZ$ mesons 
in both the $c \bar c$ and $b \bar b$ sectors,
with references to all the experiments,
are given in Ref.~\cite{Bodwin:2013nua}.
The list of new neutral $c \bar c$ mesons above the 
$D \bar D$ threshold consists of 15 states.
The most essential information in that list, including the mass, width, 
$J^{PC}$ quantum numbers, and decay modes, is repeated in 
Table~\ref{tab:ccneutral}.
This list includes an additional state labelled $Y(4220)$.
The $Y(4220)$ is a narrow structure in the cross section
for $e^+ e^-$ annihilation into $h_c(1P)\, \pi^+ \pi^-$
that was recently observed by the BESIII collaboration \cite{Yuan:2013ffw}.
Table~\ref{tab:ccneutral} also includes the additional decay mode $Z_c(3900)\, \pi$ 
of the $Y(4260)$ observed by the Belle Collaboration \cite{Liu:2013dau}.

\begin{table}[t]
\begin{center}
\begin{tabular}{cllcll}
 State & $M$~(MeV) & $\Gamma$~(MeV) & ~$J^{PC}$~ & Decay modes~ & 
     1$^{\rm st}$~observation \\
\hline
~$Z_c^+(3885)$~ & $3883.9\pm 4.5$~ & $24.8 \pm 11.5$~ & $1^{+?}$ &
     $D^{*+} \bar D^0$, $D^{+} \bar D^{*0}$ &
     BESIII~2013\\
$Z_c^+(3900)$ & $3898\pm 5$ & $51\pm 19$ & $?^{?-}$ &
     $J/\psi\, \pi^+$ &
     BESIII~2013\\
$Z_c^+(4020)$ & $4022.9\pm 2.8$~ & $7.9\pm 3.7$~ & $?^{?-}$ &
     $h_c(1P)\, \pi^+$, $D^{*+} \bar D^{*0}$~ &
     BESIII~2013\\
$Z_1^+(4050)$ & $4051^{+24}_{-43}$ & $82^{+51}_{-55}$ & $?^{?+}$ &
     $\chi_{c1}(1P)\, \pi^+$ &
     Belle~2008 \\
$Z_2^+(4250)$ & $4248^{+185}_{-\ 45}$ & 177$^{+321}_{-\ 72}$ & $?^{?+}$ &
     $\chi_{c1}(1P)\, \pi^+$ &
     Belle~2008 \\
$Z^+(4430)$ & $4443^{+24}_{-18}$ & $107^{+113}_{-\ 71}$ & $1^{+-}$ &
     $\psi(2S)\, \pi^+$ &
     Belle~2007 \\
\hline
\end{tabular}
\end{center}
\caption{Positively charged $c\bar{c}$ mesons.
The $C$ in $J^{PC}$ is that of a neutral isospin partner.
} 
\label{tab:cccharged}
\end{table}

The list of the charged $c \bar c$ mesons in Ref.~\cite{Bodwin:2013nua}
consists of 4 states.
The most essential information in that list is repeated 
in Table~\ref{tab:cccharged}.
The $C$ in $J^{PC}$
is the charge conjugation quantum number of the neutral isospin partner.
It coincides with $-G$, the negative of the $G$-parity quantum number
for the isospin triplet.
Table~\ref{tab:cccharged} gives the $J^P$ quantum numbers of the 
$Z^+(4430)$, which were recently determined to be $1^+$ by the 
Belle Collaboration \cite{Chilikin:2013tch}.
Table~\ref{tab:cccharged} includes two additional states that were observed 
more recently by the BESIII collaboration.
The $Z_c^+(3885)$ was observed in the decay channels  
$D^{*+} \bar D^0$ and $D^{+} \bar D^{*0}$, and its $J^P$ quantum numbers 
are favored to be $1^+$ \cite{Ablikim:2013xfr}.
The $Z_c^+(4020)$ was observed in the decay channel 
$h_c(1P)\, \pi^+$ \cite{Ablikim:2013wzq}.
The $Z_c^+(4025)$ was subsequently observed in the decay channel 
$(D^* \bar D^*)^+$ with a mass consistent with that of $Z_c^+(4020)$
but with a larger width \cite{Ablikim:2013emm}.
In Table~\ref{tab:cccharged}, they are assumed to be the same state.
The neutral isospin partner $Z_c^0(3900)$ of the $Z_c^+(3900)$
has been observed \cite{Xiao:2013iha}, but it is not included 
in the list of  neutral mesons in Table~\ref{tab:ccneutral}.

The decay modes of the $c \bar c$ mesons listed in 
Tables~\ref{tab:ccneutral} and \ref{tab:cccharged}
are of four kinds:
\begin{itemize}
\item
a hadronic decay into a pair of charm  mesons, such as $D \bar D$,
or a pair of charm baryons, such as $\Lambda_c^+ \Lambda_c^-$,
\item
a hadronic transition to a lighter $c \bar c$ meson through the emission of 
light hadrons, such as a single vector meson
$\omega$ or $\phi$, a single pion, or a pair of pions,
\item
an electromagnetic transition to a lighter $c \bar c$ meson 
through the emission of a photon,
\item
an electromagnetic annihilation ``decay mode''  $(e^+ e^-)$ or $(\gamma \gamma)$,
in which the parentheses indicate that it has actually been observed as a production channel.
They provide strong constraints on the $J^{PC}$ quantum numbers:
$(e^+ e^-)$ requires $1^{--}$ and $(\gamma \gamma)$ requires either
$0^{++}$ or $2^{++}$.
\end{itemize}

\begin{table}[t]
\begin{center}
\begin{tabular}{ccccll}
 State & ~$M$~(MeV)~ & ~$\Gamma$~(MeV)~ & ~$J^{PC}$~ & Decay modes & 
     1$^{\rm st}$~observation \\
\hline
~$Y_b(10890)$~ & 10888.4$\pm$3.0 & 30.7$^{+8.9}_{-7.7}$ & $1^{--}$ &
      $\Upsilon(nS)\, \pi^+\pi^-$,  ($e^+ e^-$)~ &  Belle~2010 \\
\hline
~$Z_{b}^+(10610)$~ & ~10607.2$\pm$2.0~ & ~18.4$\pm$2.4~ & $1^{+-}$ &
       $\Upsilon(nS)\,\pi^+$, $h_b(nP)\,\pi^+$~ &
      Belle~2011 \\
 &  &  & & $\bar B^{*0}  B^+$, $\bar B^0  B^{*+}$ & \\
$Z_{b}^+(10650)$ & 10652.2$\pm$1.5 & 11.5$\pm$2.2 & $1^{+-}$ &
       $\Upsilon(nS)\,\pi^+$, $h_b(nP)\,\pi^+$ &
    Belle~2011 \\
 &  &  & & $\bar B^{*0}  B^{*+}$ & \\
\hline
\end{tabular}
\end{center}
\caption{Neutral and positively charged $b\bar{b}$ mesons above the $B\bar{B}$ threshold
discovered since 2003.  
Neutral isospin partners of charged $b \bar  b$ mesons are not listed.
For a charged $b\bar{b}$ meson,
the $C$ in $J^{PC}$ is that of the neutral isospin partner.
}
\label{tab:bbboth}
\end{table}

The list of new $b \bar b$ mesons above the $B \bar B$ threshold 
in Ref.~\cite{Bodwin:2013nua}
consists of 1 neutral and 2 charged states.
The most essential information in that list is repeated in Table~\ref{tab:bbboth}.
Table~\ref{tab:bbboth} includes additional decay modes of the 
$Z_{b}^+(10610)$ and $Z_{b}^+(10650)$ into pairs of bottom mesons \cite{Adachi:2012cx}.
The neutral isospin partner $Z_b^0(10610)$ of the $Z_b^+(10610)$
has been observed \cite{Adachi:2012im}, but it is not included 
in the list of neutral mesons in Table~\ref{tab:bbboth}.

A theoretical framework for the $XYZ$ mesons should explain the pattern of 
all the observed mesons, including their masses, widths, quantum numbers, and decay modes.
It should also predict other $XYZ$ mesons that await discovery.
The theoretical framework should be based as closely as possible 
on the fundamental field theory QCD.  
It should involve the fundamental degrees of freedom of QCD, 
which are quark and gluon fields, and any interactions should be derivable from 
the fundamental QCD interactions, which are mediated by the exchange of gluons.
The B-O approximation provides such a  theoretical framework.

\section{Born-Oppenheimer Potentials}
\label{sec:BOpotential}

In this section, we discuss the behavior of the various 
Born-Oppenheimer (B-O) potentials for $Q \bar Q$ mesons.  
We give simple analytic approximations for the deepest of the hybrid potentials
that have been calculated using lattice QCD.  
We also infer the deepest of the tetraquark potentials
from lattice QCD calculations of adjoint mesons.

\subsection{Definitions of Born-Oppenheimer potentials}
\label{sec:defBOpot}

In QCD without light quarks, the ground-state B-O potential 
$V_{\Sigma_g^+}(r)$ can be defined as the minimal energy for
configurations of the gluon field in the presence of $Q$ and $ \bar Q$ sources
separated by a distance $r$.  An excited B-O potential 
$V_\Gamma(r)$ can be defined as the minimal energy for
configurations of the gluon field with quantum numbers specified by
$\Gamma$, provided $V_\Gamma(r)$ is
smaller than the sum of $V_{\Sigma_g^+}(r)$ 
and the mass of a glueball with the appropriate quantum numbers.    
Otherwise, the minimal-energy configuration is the $\Sigma_g^+$
gluon configuration accompanied by a zero-momentum glueball.  
A potential $V_\Gamma(r)$ that is larger than the sum of $V_{\Sigma_g^+}(r)$ 
and the mass of the glueball may still be well-defined 
as the energy of a stationary gluon configuration that is localized 
near the line connecting the $Q$ and $\bar Q$. 
A prescription for the potential might
involve calculating the energies of excited configurations of the
gluon field with quantum numbers $\Gamma$ 
in the presence of $Q$ and $\bar Q$ sources separated by $r$,
and identifying the potential $V_\Gamma(r)$
as the energy of one of the excited configurations.

In QCD with light quarks, there are additional complications 
in the definitions of the excited flavor-singlet B-O potentials.  
For small $r$, the minimal-energy configuration is the 
$\Sigma_g^+$ configuration accompanied by either two or three pions,
depending on the quantum numbers of $\Gamma$.   
For large $r$, the minimal-energy configuration consists of 
two {\it static mesons}, which are configurations of the light-quark and gluon fields
bound to a static $Q$ or $\bar Q$ source.
One of the static mesons has the flavor
of a light antiquark $\bar q$ and is localized near the $Q$ source,
while the other has the flavor of a light quark $q$ 
land is localized near the $\bar Q$ source. 
The energy of such a configuration defines a B-O potential
that approaches a constant as $r \to \infty$.
As $r$ decreases, the extrapolation of this potential crosses the  
extrapolations of the $\Sigma_g^+$ and other potentials.
However there are actually avoided crossings
between pairs of potentials that share the same quantum numbers $\Gamma$.
If $r$ is not too close to an avoided crossing,
a B-O potential $V_\Gamma(r)$ may still be well-defined 
as the energy of a stationary configuration of the gluon and light-quark fields. 
A prescription for the potential might
involve calculating the energies of excited configurations of 
gluon and light-quark fields with quantum numbers $\Gamma$ 
in the presence of $Q$ and $\bar Q$ sources separated by $r$,
and identifying the potential $V_\Gamma(r)$ 
as the energy of one of the excited configurations.

Light quarks introduce an additional complication 
that is the key to understanding the tetraquark $XYZ$ states. 
The gluon and light-quark configurations in the presence of 
static $Q$ and $\bar Q$ sources are specified not only by the 
traditional quantum numbers $\Gamma$ of the Born-Oppenheimer (B-O) 
approximation but also by light-quark flavor quantum numbers. 
As pointed out in Ref.~\cite{Braaten:2013boa}, 
B-O potentials can also be defined for isospin-1 configurations 
of light-quark and gluon fields. 
They can also be defined for isospin-0 configurations 
and for configurations that contain a strange quark and a lighter antiquark.
The energy levels in these 
potentials are tetraquark mesons. Thus the B-O 
approximation can be used to describe conventional quarkonium, 
quarkonium hybrids, and quarkonium tetraquarks all 
within a common framework. The definition of flavor-nonsinglet 
tetraquark potentials suffers from the same complications 
as the excited-state flavor-singlet B-O potentials. 
At large $r$, the minimal-energy configuration consists of 
two static mesons localized near the $Q$ and $\bar Q$ sources.
At small $r$, the minimal-energy configuration is the 
flavor-singlet $\Sigma_g^+$ potential accompanied by one or two pions,
depending on the quantum numbers $\Gamma$.  
Thus the minimal-energy prescription is inadequate,
and it is necessary to use a more complicated prescription
to define the flavor-nonsinglet B-O potentials.  
A prescription for the potential might
involve calculating the energies of excited configurations of 
gluon and light-quark fields with quantum numbers $\Gamma$ 
and the appropriate flavors
in the presence of $Q$ and $\bar Q$ sources separated by $r$,
and identifying the potential $V_\Gamma(r)$ 
as the energy of one of the excited configurations.

\subsection{Light-field quantum numbers}
\label{sec:lightQN}

The B-O potentials can be labelled by quantum numbers for the 
gluon and light-quark fields that are conserved  
in the presence of static $Q$ and $ \bar Q$ sources.  
We first consider the flavor-singlet case. Let $\bm{r}$ be 
the separation vector between the $Q$ and $\bar Q$ sources.  
There are three conserved quantum numbers for the light fields
in the presence of these sources:
\begin{itemize}
\item 
the eigenvalue $\lambda$ of
$\hat{\bm{r}}\cdot \bm{J}_{\rm light}$, where $\bm{J}_{\rm light}$ 
is the total angular momentum vector for the light fields.  
The possible values of $\lambda$ are 
0, $\pm 1$, $\pm 2$, \ldots.  
We denote its absolute value by $\Lambda$:   $\Lambda = |\lambda|$.

\item 
the eigenvalue $\eta$ of 
$(CP)_{\rm light}$, which is the product of the charge-conjugation 
operator $C_{\rm light}$ for the light fields and the 
parity operator $P_{\rm light}$ that spacially inverts these fields through 
the midpoint between the $Q$ and $\bar Q$ sources.  
The possible values of $\eta$ are $+1$ and $-1$.  

\item 
for the  case $\lambda = 0$, the eigenvalue $\epsilon$
of a reflection operator $R_{\rm light}$ that reflects the light fields
through  a plane containing the $Q$ and $\bar Q$ sources.
The possible values of $\epsilon$ are $+1$ or $-1$. 
\end{itemize}
It is traditional to use an upper-case 
Greek letter to specify the integer $\Lambda$: 
$\Sigma$ for $\Lambda = 0$, $\Pi$ for $\Lambda = 1$, 
$\Delta$ for $\Lambda =  2$, etc.   
The eigenvalue $+1$ or $-1$ of $(CP)_{\rm light}$ is traditionally specified 
by a subscript $g$ or $u$ on the upper-case Greek letter.  
In the case $\lambda = 0$, the value $+1$ or $-1$ of $\epsilon$ 
is traditionally specified 
by a superscript $+$ or $-$ on $\Sigma$.
Thus the B-O potentials are traditionally labelled 
by $\Gamma = \Sigma_\eta^+,\Sigma_\eta^-,\Pi_\eta, \Delta_\eta, \ldots$, 
where the subscript $\eta$ is $g$ or $u$.

The stationary configuration of gluon and light-quark fields
associated with the quantum numbers
$\lambda=0$, $\eta$, and $ \epsilon$ can be represented by the ket
$| 0, \eta, \epsilon\, ; \bm{r} \rangle$. It is an eigenstate
of $(CP)_{\rm light}$ and $P_{\rm light}$ with eigenvalues $\eta$
and $\epsilon$, respectively.
The stationary configuration of gluon and light-quark fields
associated with the quantum numbers
$\lambda\neq 0$ and $\eta$ can be represented by the ket
$| \lambda, \eta; \bm{r} \rangle$.
The reflection operator $R_{\rm light}$ can be expressed as the product 
of the parity operator $P_{\rm light}$ and a rotation by angle
$\pi$ around the axis perpendicular to the reflection plane 
and passing through the midpoint between the $Q$ and $\bar Q$ sources.  
It maps a configuration $| \lambda, \eta; \bm{r} \rangle$
with $|\lambda| \ge 1$ into $(-1)^\lambda | -\lambda, \eta; \bm{r} \rangle$.
The reflection symmetry guarantees that the configurations
$| \Lambda, \eta; \bm{r} \rangle$ and $| -\Lambda, \eta; \bm{r} \rangle$ 
have the same energies.
We can form linear combinations of these states that are eigenstates
of $R_{\rm light}$:
\begin{equation}
\big| \Lambda, \eta, \epsilon\, ; \bm{r} \big\rangle \equiv \frac{1}{\sqrt{2}}
\Big( \big| \Lambda, \eta ; \bm{r} \big\rangle 
+ \epsilon\; \big| -\Lambda, \eta; \bm{r} \big\rangle \Big).
\label{Gamma_eta^pm}
\end{equation}	
They are eigenstates of $|\hat{\bm{r}}\cdot \bm{J}_{\rm light}|$,	
$(CP)_{\rm light}$, and $R_{\rm light}$ with eigenvalues 
$\Lambda$, $\eta$, and $ \epsilon$, where $\epsilon$ is $+1$ or $-1$.
They are also eigenstates of $P_{\rm light}$ with eigenvalue $\epsilon\,  (-1)^\Lambda$.
Thus the stationary light-field configurations 
can be labeled by the quantum numbers
$ \Lambda$, $\eta$, and $ \epsilon$ or alternatively by
$\Gamma = \Sigma_\eta^\epsilon, \Pi_\eta^\epsilon, \Delta_\eta^\epsilon,
\ldots$, where the subscript $\eta$ is $g$ or $u$ and 
the superscript $\epsilon$ is $+$ or $-$.

For quarkonium tetraquark mesons, the stationary configurations of the gluon 
and light-quark fields also have flavor quantum numbers.  
The flavor quantum numbers can be identified 
by specifying the light quark and antiquark: 
$q_1 \bar q_2$, where $q_1,q_2=u,d,s$.  
We proceed to discuss the conserved quantum numbers for light-field
configurations with flavor quantum numbers $q_1 \bar q_2$
in the presence of static $Q$ and $\bar Q$ sources.
The eigenvalue $\lambda$ of $\hat{\bm{r}}\cdot \bm{J}_{\rm light}$
remains conserved.  If $q_1$ and $q_2$ are distinct flavors,
$C_{\rm light}$ changes the flavor from $q_1 \bar q_2$
to $q_2 \bar q_1$.  Thus $(CP)_{\rm light}$ 
followed by the flavor interchange  $q_1 \leftrightarrow q_2$  
is a symmetry of the light-field configurations. 
Its quantum number $\eta$ is conserved.
If $\lambda=0$, the reflection quantum number $\epsilon$ is also conserved.
Thus the stationary field configurations can be labelled by
$q_1 \bar q_2$, $\lambda$, $\eta$, and also $\epsilon$ if $\lambda=0$.
Alternatively, for $|\lambda| \ge 1$, we can form linear combinations
$| q_1 \bar q_2; \Lambda, \eta, \epsilon\,;  \bm{r}  \rangle$ 
analogous to Eq.~(\ref{Gamma_eta^pm}) that are labelled
by $\Lambda= |\lambda|$, $\eta$, and $\epsilon$.
These configurations are eigenstates of  $R_{\rm light}$
with the eigenvalue $\epsilon$.

The tetraquark potentials with flavor $q_1 \bar q_2$ are energies of
stationary configurations of light-quark and gluon fields.
For $q_1,q_2=u,d$, it is more convenient to use the 
isospin quantum numbers $(I,I_3)$ to specify the flavor state: 
$(0,0) = (u \bar u + d \bar d)/\sqrt{2}$, $(1,+1) = -u \bar d$, 
$(1,0) = (u \bar u - d \bar d)/\sqrt{2}$, and $(1,-1)= d \bar u$.
If we ignore small effects from the 
difference between the masses of the $u$ and $d$ quarks and their    
different electromagnetic charges, 
the energy of the configuration 
depends on the isospin quantum number $I$, which is 0 or 1, but not on $I_3$. 
Within the same approximation, 
the flavor configurations $u \bar s$ and $d\bar s$, which form an isospin doublet,
and the flavor configurations $-s \bar d$ and $s \bar u$, 
which also form an isospin doublet,
all have the same energies.  Thus the only flavor labels required to specify the 
distinct energies of the tetraquark configurations are $I=0$, $I=1$, 
$s \bar q$, and $s \bar s$.
The tetraquark potentials
can be denoted $V_\Gamma^{(I=0)}(r)$, $V_\Gamma^{(I=1)}(r)$, 
$V_\Gamma^{(s \bar q)}(r)$, and $V_\Gamma^{(s \bar s)}(r)$,
where $\Gamma = \Lambda_\eta^\epsilon$.

If we consider only flavor quantum numbers with no net strangeness 
(i.e.\ either $q_1\bar q_2$ with $q_1,q_2= u,d$ or else $s \bar s$), 
the stationary light-field configurations with flavors $I=0$, $I=1$, 
and $s \bar s$
are also eigenstates of $G$-parity, which is the product of the charge conjugation operator
$C_{\rm light}$ and an isospin rotation by angle $\pi$ around the $I_2$ axis.
Its eigenvalues are:
\begin{equation}
G =  \eta\,  \epsilon\,  (-1)^{\Lambda+I}.
\label{Glight}
\end{equation}	
The neutral members of the isospin multiplets are eigenstates of $C_{\rm light}$
with eigenvalue $-G$.

\subsection{Quarkonium potential}
\label{sec:oniumpot}

Conventional quarkonia are energy levels of a $Q \bar Q$ pair
in the flavor-singlet $\Sigma_g^+$ potential. 
The limiting behaviors of the $\Sigma_g^+$ potential 
at large $r$ and at small $r$ are understood,
at least in the absence of light quarks \cite{Juge:2002br}. 
At large $r$, the field configuration for $\Sigma_g^+$
is a flux tube extending between the $\bar Q$ and $Q$. 
The potential approaches the ground-state energy of a 
relativistic string of length $r$ with fixed endpoints  \cite{Juge:2002br}:
\begin{equation}
V_{\Sigma_g^+}(r) \longrightarrow 
\sigma r \left(1 - \frac{\pi}{6\sigma r^2} \right)^{1/2} + E_0,
\label{VSigmag+:largeR}
\end{equation}	
where $\sigma$ is the string tension, which is the energy per length of the flux tube,
and $E_0$ is an additive constant. 
At small $r$, 
the $\Sigma_g^+$ potential approaches the attractive color-Coulomb potential 
between a $Q$ and $\bar Q$ in a color-singlet state: 
\begin{equation}
V_{\Sigma_g^+}(r) \longrightarrow 
- \frac{4\alpha_s(1/r)}{3r} + E_{\Sigma_g^+},
\label{VSigmag+:smallR}
\end{equation}	
where $\alpha_s(\mu)$ is the running coupling constant of QCD 
at the momentum scale $\mu$ and $E_{\Sigma_g^+}$ is an additive constant \cite{Juge:2002br}. 

A simple phenomenological potential that is qualitatively 
compatible with the limiting behaviors in Eqs.~(\ref{VSigmag+:largeR})
and (\ref{VSigmag+:smallR})
is the Cornell potential \cite{Eichten:1974af}:
\begin{equation}
V_{\Sigma_g^+}(r) 
= 2 m_Q + V_0 - \frac{\kappa}{r} + \sigma r .
\label{VCornell-R}
\end{equation}	
The parameter $\kappa$ 
can be interpreted as an effective value of $(4/3)\alpha_s (1/r)$ 
in the small-$r$ region.
Alternatively, if the value of $\kappa$ is close to $\pi/12 \approx 0.262$, 
it can be interpreted as a coefficient in the expansion of 
Eq.~(\ref{VSigmag+:largeR}) at large $r$.
The additive constant in Eq.~(\ref{VCornell-R}) has been separated into $2 m_Q$
and a term $V_0$ that is independent of the heavy quark. 
The parameters $\sigma$,   $\kappa$, $m_c$, $m_b$ and 
$V_0$  can all  be determined 
phenomenologically by fitting the energy levels of conventional 
charmonium and bottomonium.
Such a fit will be carried out in Section~\ref{sec:pheno-onium}.

\begin{figure}[t]
\centerline{ \includegraphics*[width=16cm,clip=true]{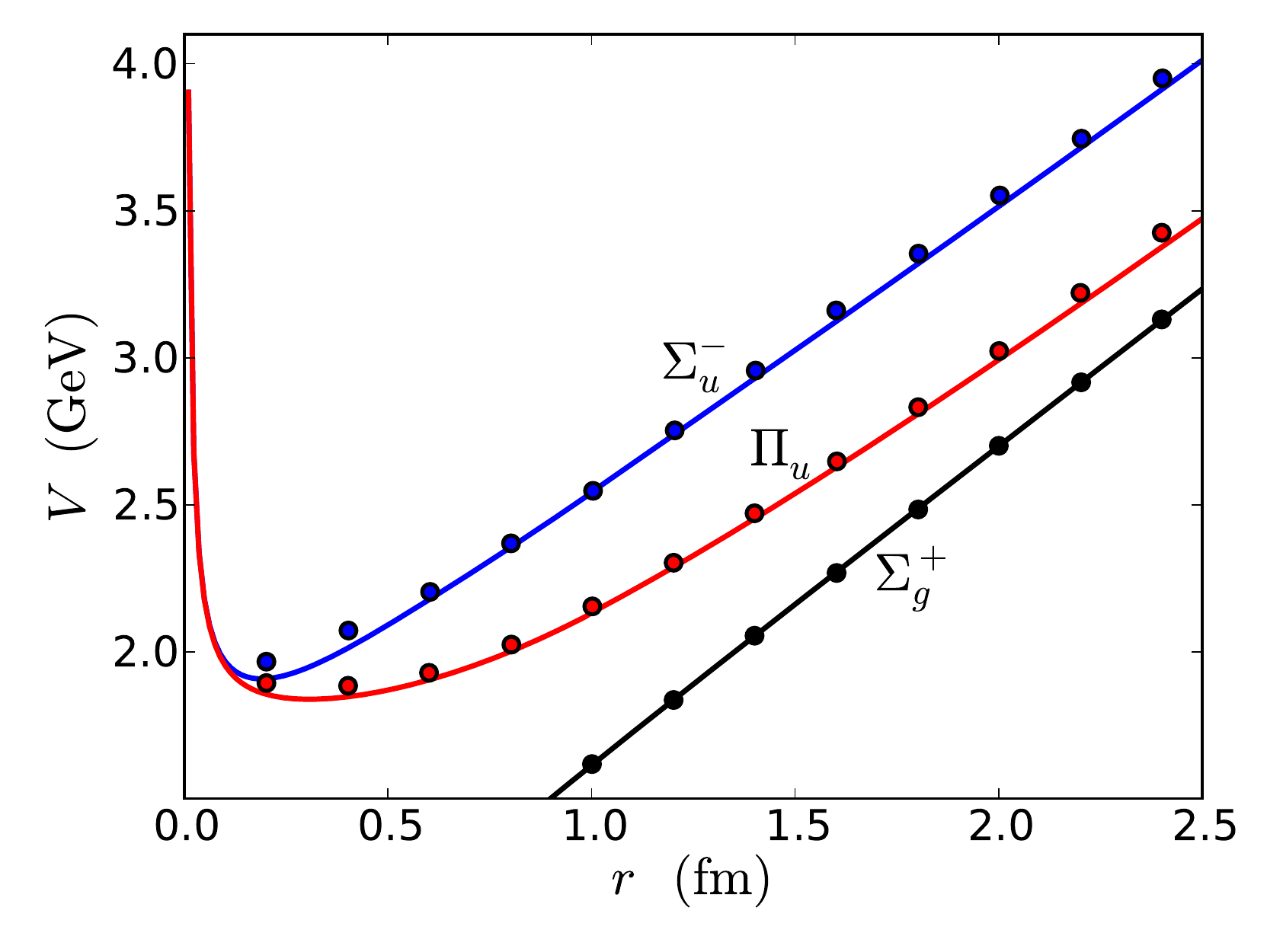} }
\vspace*{0.0cm}
\caption{The $\Sigma_u^+$, $\Pi_u$, and $\Sigma_u^-$ Born-Oppenheimer potentials for 
conventional quarkonium and quarkonium hybrids.
The dots are the potentials calculated using quenched lattice gauge theory in Ref.~\cite{Juge:1999ie}.
The curves are fits to those potentials.}
\label{fig:Vlattice}
\end{figure}

The $\Sigma_g^+$ potential can be calculated using lattice QCD.
The string tension $\sigma$ defined by Eq.~(\ref{VSigmag+:largeR})
can be used to set the length scale in lattice QCD calculations.
However calculations of potentials in lattice QCD are more stable
if the length scale is set instead by the Sommer radius 
$r_0$   \cite{Sommer:1993ce} defined by 
\begin{equation}
r_0^2V'_{\Sigma_g^+}(r_0) = 1.65,
\label{Sommer}
\end{equation}	
where $V'(r)$ represents the derivative of the potential with respect to $r$.
A phenomenological value of this parameter obtained by fitting the 
bottomonium spectrum is $r_0^{-1} = 394\pm20$~MeV \cite{Bali:2000gf},
so $r_0 \approx 0.50$~fm.  
A fit of the $\Sigma_g^+$ potential calculated using quenched lattice QCD
to the Cornell potential gives $\kappa = 0.292(6)$ and 
$\sqrt{\sigma} r_0 = 1.165(3)$   \cite{Bali:2000vr},
which implies $\sigma = 0.21 \pm 0.02~{\rm GeV}^2$.
The long-distance part of the potential calculated in Ref.~\cite{Juge:1999ie} 
using quenched lattice gauge theory is shown in Fig.~\ref{fig:Vlattice}.
If the potential is fit to Eq.~(\ref{VSigmag+:largeR}),
the string tension is determined to be $\sigma = 0.21~{\rm GeV}^2$.

The $\Sigma_g^+$ potential has been calculated using lattice QCD
with two flavors of dynamical light quarks \cite{Bali:2000vr}.  
A fit to the Cornell potential gives $\kappa = 0.368^{+20}_{-26}$
and $\sqrt{\sigma} r_0 = 1.133^{+11}_{-8}$,
which implies $\sigma = 0.20\pm0.02~{\rm GeV}^2$.
The values of $\kappa$ and $\sqrt{\sigma} r_0$ differ significantly from
those calculated using quenched lattice QCD.
The extrapolation of the $\Sigma_g^+$ potential crosses the threshold 
defined by twice the mass of the static meson near $r \approx 2.4\, r_0$.  
For $r > 2.4\, r_0$, the minimal-energy prescription can not be used to define 
the $\Sigma_g^+$ potential.  
Lattice QCD has nevertheless been used to calculate the $\Sigma_g^+$ potential 
with high precision at larger values of $r$ \cite{Bolder:2000un}.
There is another B-O potential that approaches the threshold for a pair of static mesons
as $r \to \infty$.  It actually has an avoided crossing with the $\Sigma_g^+$ potential.
Lattice QCD has been used to calculate the two potentials 
in the region of the avoided crossing \cite{Bali:2005fu}.

\subsection{Hybrid potentials}
\label{sec:hybridpot}

Quarkonium hybrid mesons are energy levels  of a $Q \bar Q$ pair
in the excited flavor-singlet B-O potentials.  
We will refer to these potentials as hybrid potentials. 
Many of the hybrid potentials were calculated by Juge, Kuti, 
and Morningstar using quenched lattice QCD \cite{Juge:1999ie,Juge:2002br}.  
They all have minima at positive values of $r$.
The deepest hybrid potentials are $\Pi_u$ and $\Sigma_u^-$.  

The limiting behaviors of the hybrid potentials 
at large $r$ and at small $r$ are understood,
at least in the absence of light quarks  \cite{Juge:2002br}. 
At large $r$, the field configuration for the potential $\Gamma$ 
is a flux tube extending between the $Q$ and $\bar Q$.
The corresponding potential
approaches an excited energy level of a relativistic 
string of length $r$ with fixed endpoints:
\begin{equation}
V_\Gamma(r)\longrightarrow
\sigma r \left(1+\frac{\pi(12 n_\Gamma-1)}{6 \sigma r^2}\right)^{1/2} + E_0,
\label{VGamma:largeR}
\end{equation}	
where the excitation number $n_\Gamma$ depends on the B-O potential,
and $E_0$ is the same additive constant as in Eq.~(\ref{VSigmag+:largeR}). 
The $\Sigma_g^+$ potential, whose limiting behavior is given in
Eq.~(\ref{VSigmag+:largeR}), 
is the ground state of the string with $n_\Gamma = 0$. 
The $\Pi_u$ potential is the first excited state with
$n_\Gamma = 1$.
The $\Sigma_u^-$ potential has $n_\Gamma = 3$.
The B-O potentials with  $n_\Gamma = 2$ are $\Delta_g$,
$\Pi_g$, and the first excited $\Sigma_g^+$ potential,
which is labelled $\Sigma_g^+{}'$.
At small $r$, the hybrid potentials approach the repulsive 
color-Coulomb potential between a $Q$ and $\bar Q$ in a color-octet state: 
\begin{equation}
V_{\Gamma}(r) \longrightarrow
+ \frac{ \alpha_s(1/r)}{6r} + E_\Gamma,
\label{VGamma:smallR}
\end{equation}	
where $E_\Gamma$ is an additive constant that depends on the B-O potential.

In the limit $r \to 0$, the $Q$ and $\bar Q$ sources reduce to a 
single local color-octet $Q \bar Q$ source.
In this limit, the conserved quantum numbers of the gluon 
and light-quark fields in the presence of the source
are $J^{PC}_{\rm light}$.
The energy levels of flavor-singlet gluon and light-quark field configurations
bound to a static color-octet source are called
{\it static hybrid mesons} or {\it gluelumps}. 
In QCD without light quarks, a gluelump can be defined as the 
minimal-energy configuration of the gluon field with specified quantum numbers
$J^{PC}_{\rm light}$.
In QCD with light quarks, the minimal-energy prescription 
can still be used to define the ground-state gluelump 
with quantum numbers $1^{+-}$.
The minimal-energy prescription 
can be used to define an excited gluelump 
only if its energy relative to the ground-state gluelump is less than $2 m_\pi$
or $3 m_\pi$, depending on the quantum numbers.
Otherwise, the minimal-energy configuration
is the ground-state $1^{+-}$ gluelump accompanied by 2 or 3 pions.
In this case, the excited gluelump 
would have to be identified as  one of the
excited states of the gluon and light-quark fields
with the appropriate $J^{PC}_{\rm light}$ quantum numbers
in the presence of the static $Q \bar Q$ source.

In the limit $r \to 0$, the gluon and light-quark fields in the presence 
of the $Q$ and $\bar Q$ sources have additional symmetries
that require various B-O potentials 
to become degenerate in that limit \cite{Brambilla:1999xf}.
In the limiting expression  in Eq.~(\ref{VGamma:smallR}) 
for the hybrid potential at small $r$, 
the additive constant $E_\Gamma$ can be interpreted as the energy of a gluelump.  
For the two deepest hybrid potentials,
$\Pi_u$ and $\Sigma_u^-$, $E_\Gamma$ must be equal
to the energy of the ground-state $1^{+-}$ gluelump.
For the $\Pi_g$ and $\Sigma_g^+{}'$ potentials, $E_\Gamma$ 
must be equal to the energy of the $1^{--}$ gluelump.
For the $\Delta_g$ potential, the  $\Sigma_g^-$ potential,
and the first excited $\Pi_g$ potential, which is labelled $\Pi'_g$, 
$E_\Gamma$ must be equal to the energy of the $2^{--}$ gluelump.

Given the quantum numbers $J^{PC}_{\rm light}$ of the gluelump,
we can deduce the hybrid potentials whose additive constant 
$E_\Gamma$ defined by Eq.~(\ref{VGamma:smallR})
is equal to the energy of the gluelump.
A component of the angular momentum vector for a gluelump with spin
$J_{\rm light}$ has $2J_{\rm light} + 1$ integer values ranging
from $-J_{\rm light}$ to $+J_{\rm light}$.  There must therefore be a
B-O potential for each integer value of $\Lambda$ from 0 up to $J_{\rm light}$.
The quantum number $\eta$ for all these potentials equals the value of 
$(CP)_{\rm light}$ for the gluelump.
One of the B-O potentials must be a $\Sigma$ potential with $\Lambda = 0$.
If we identify its reflection quantum number as 
$\epsilon = P_{\rm light} (-1)^{J_{\rm light}}$,
the $\Sigma$ potentials associated with the $1^{+-}$, $1^{--}$, and $2^{--}$ gluelumps
are correctly inferred to be $\Sigma_u^-$, $\Sigma_g^+{}'$, and $\Sigma_g^-$,
respectively.

\begin{table}[t]
\begin{center}
\begin{tabular}{cc|ccc}
\multicolumn{2}{c}{gluelumps} & \multicolumn{3}{c}{adjoint mesons} \\
~$J^{PC}$~ & ~$g$~ & ~$J^{PC}$~ & ~$q \bar q$~ &~$s \bar s$~  \\
\hline
$1^{+-}$ &             (0)           & $1^{--}$ & ~$47 \pm 90~$~ & ~$120 \pm 70$~  \\
$1^{--}$  & ~$285 \pm 53$~ & $0^{-+}$ & ~$91 \pm 216$ & ~$170 \pm 99$~ \\
$2^{--}$  & ~$710 \pm 37$~ &             &                          &                               \\
\hline
\end{tabular}
\end{center}
\caption{Gluelump and adjoint meson energies (in MeV) 
relative to the energy of the $1^{+-}$ ground-state gluelump.
The gluelump energies are from the lattice QCD calculations 
with dynamical light quarks in Ref.~\cite{Marsh:2013xsa}.
The adjoint meson energies are from the quenched lattice QCD calculations 
in Ref.~\cite{Campbell:1985kp}.
The errors in the adjoint meson energies do not take into account
systematic errors from omitting light-quark loops.}
\label{tab:gl,am}
\end{table}

The gluelump spectrum was first calculated 
using quenched lattice QCD by Campbell, 
Jorysz, and Michael \cite{Campbell:1985kp}. 
The ground-state gluelump
was found to have quantum numbers $1^{+-}$. 
More accurate results for the 
gluelump energy differences were calculated subsequently by 
Foster and Michael \cite{Foster:1998wu}. 
The gluelump spectrum was recently calculated by Marsh and Lewis 
using lattice QCD with dynamical light quarks \cite{Marsh:2013xsa}. 
The strange quark had its physical mass, but the up and down 
quark masses were unphysically heavy, 
corresponding to a pion mass of about 500 MeV. 
The first two excited states of the gluelump have quantum numbers $1^{--}$
and $2^{--}$.  Their energies relative to that of the ground-state $1^{+-}$ gluelump
are given in Table~\ref{tab:gl,am}.
The energies of the $1^{--}$ and $2^{--}$ gluelumps are higher
by about 300~MeV and  700~MeV, respectively.

The hybrid potentials can be calculated using lattice QCD.
In Ref.~\cite{Juge:1999ie}, quenched lattice QCD  
was used to calculate many of these potentials.
They interpolate between the short-distance limit in Eq.~(\ref{VGamma:smallR}) 
and the long-distance limit in Eq.~(\ref{VGamma:largeR}). 
The $\Pi_u$ and $\Sigma_u^-$ potentials from Ref.~\cite{Juge:1999ie}
are shown in Fig.~\ref{fig:Vlattice}.

An accurate parametrization of the $\Pi_u$ potential in 
Ref.~\cite{Juge:1999ie} at short and intermediate distances 
was given in Ref.~\cite{Vairo:2009tn}:
\begin{equation}
V_{\Pi_u}(r) = E_{\Pi_u} + 0.11\frac{1}{r} + 0.24\frac{r^2}{r_0^3},
\label{VPiu:intR}
\end{equation}	
where $r_0 \approx 0.50$~fm is the Sommer radius defined in Eq.~(\ref{Sommer}).
This parametrization provides an excellent fit to the $\Pi_u$ potential
at lattice spacings between $0.3\, r_0$ and $2.4\, r_0$.
The minimum of the potential is at $0.61 r_0 \approx 0.3$~fm.
The $\Pi_u$ potential has also been calculated for QCD with
two flavors of dynamical light quarks \cite{Bali:2000vr}.
No statistically significant differences were found between the 
$\Pi_u$ potentials with and without light quarks.

An important feature of the parametrization of the $\Pi_u$ potential
in Eq.~(\ref{VPiu:intR}) is the absence of a linear term in $r$.
The $1/r$ term can be interpreted as the repulsive color-Coulomb potential 
between the $Q$ and $\bar Q$.  The remaining terms can be 
interpreted as the energy of the gluon field configuration.
According to the parametrization in Eq.~(\ref{VPiu:intR}),
the gluon field energy has zero slope at $r=0$, 
so it increases slowly with $r$ in the small-$r$ region.
At $r=0$, the gluon field configuration is the ground-state gluelump.
At $r=r_0$, the gluon field energy has increased by less than 1/3 of the
energy difference for the first excited gluelump.
The slow increase of the gluon field energy with $r$ is consistent
with  the gluon field configuration remaining close to the gluelump
out to values of $r$ comparable to $r_0$.  
At these small values of $r$, the configuration is compatible with the 
simple constituent model of a quarkonium hybrid meson: $(Q \bar Q)_8 + g$.
The constituent gluon $g$ can be identified with the ground-state 
gluelump with quantum numbers $1^{+-}$.

We can obtain a global fit to the potential $V_{\Pi_u}(r)$
by using the parametrization in Eq.~(\ref{VPiu:intR}) for $r$ 
below some matching radius $r_*$ and then switching to the 
string potential in Eq.~(\ref{VGamma:largeR}) with $n_\Gamma = 1$
for $r$ beyond $r_*$. We demand continuity of the potentials and their slopes 
at the matching point $r_*$:
\begin{subequations}
\begin{eqnarray}
E_{\Pi_u} - E_0 + 0.11 \frac{1}{r_*} + 0.24 \frac{r_*^2}{r_0^3} &=& 
\sigma \sqrt{r_*^2 + 11 \pi/(6 \sigma)},
\label{VPimatch1}
\\
- 0.11\frac{1}{r_*} + 0.48 \frac{r_*^2}{r_0^3} &=& 
\frac{\sigma r_*^2}{\sqrt{r_*^2 + 11 \pi/(6 \sigma)}}.
\label{VPimatch2}
\end{eqnarray}	
\label{VPimatch}%
\end{subequations}
These two equattions determine $r_*$ and $E_{\Pi_u} - E_0$.
If we take the value $\sqrt{\sigma} = 0.21~r_0^{-1}$ from the fit to the 
long-distance part of the 
$\Sigma_g^+$ potential in Ref.~\cite{Juge:2002br}, which is shown in Fig.~\ref{fig:Vlattice}, 
the matching point is determined by Eq.~(\ref{VPimatch2})
to be $r_* = 2.0\, r_0$. The difference between the energy offsets
is then determined by Eq.~(\ref{VPimatch1}) to be  $E_{\Pi_u} - E_0 = 2.8\, r_0^{-1}$.  
In Fig.~\ref{fig:Vlattice}, the resulting parametrization of
$V_{\Pi_u}(r)$ is compared to the potential calculated using 
quenched lattice QCD in Ref.~\cite{Juge:2002br}.
It gives a good fit over the entire range of $r$.
The fit could be slightly improved by relaxing the constraint that the long-distance limits 
of the ground-state $\Sigma_g^+$ potential in Eq.~(\ref{VSigmag+:largeR}) 
and the excited-state potentials  $\Gamma$ in Eq.~(\ref{VGamma:largeR}) 
have the same additive constant $E_0$.

To obtain a parametrization of the $\Sigma_u^-$ potential,
it is most convenient to fit the difference between the 
$\Sigma_u^-$ and $\Pi_u$ potentials.
In the quenched lattice QCD calculations in Ref.~\cite{Juge:2002br},
that difference appears to be linear in $r$ at small $r$.
In Ref.~\cite{Bali:2003jq}, the splitting between
$V_{\Sigma_u^-}(r)$ and $V_{\Pi_u}(r)$ was calculated in the region $r < 1.6~r_0$ 
using quenched lattice QCD with a finer lattice.  
The results are consistent with those of Ref.~\cite{Juge:2002br}, 
but they extend down to smaller values of $r$.  For $r< 0.8~ r_0$,
the splitting is compatible with quadratic dependence on $r$,
and it can be fit with $0.92~r^2/r_0^3$.
Given the constraint provided by this leading power of $r^2$,
the difference between the potentials in Ref.~\cite{Juge:2002br}
can be fit very well with the simple parametrization
\begin{equation}
V_{\Sigma_u^-}(r) = V_{\Pi_u}(r) 
+ \frac{0.92~r^2/r_0^3} {1 + 0.63~r^2/r_0^2}.
\label{VSigmau:intR}
\end{equation}	
The minimum of the $\Sigma_u^-$ potential is near $0.4~r_0 \approx 0.2$~fm.
In Fig.~\ref{fig:Vlattice}, the parametrization of
$V_{\Sigma_u^-}(r)$ is compared to the potential calculated using 
quenched lattice QCD in Ref.~\cite{Juge:2002br}.
It gives a good fit over the entire range of $r$.
It will not give a good fit for $r> 2.5$~fm, because the parametrization
in Eq.~(\ref{VSigmau:intR}) does not take into account the constraints	
from the large-$r$ limit given by Eq.~(\ref{VGamma:largeR}).

\subsection{Tetraquark potentials}
\label{sec:tetrapot}

Quarkonium tetraquark mesons are energy levels in B-O potentials 
with nonsinglet flavor quantum numbers, such as $q_1 \bar q_2$. 
We will refer to these potentials as tetraquark potentials. 
The distinct B-O potentials can be specified by the flavor labels
$I=0$, $I=1$, $s \bar q$, and $s \bar s$ and by the quantum numbers 
$\Gamma = \Sigma_\eta^+,\Sigma_\eta^-,\Pi_\eta,\Delta_\eta,\ldots$ for the 
light-quark and gluon field  configuration.
None of the tetraquark potentials have yet been calculated using lattice QCD. 

The only information about the tetraquark potentials that is known from lattice QCD
comes from calculations of {\it static adjoint mesons}, which are energy levels 
of light-quark and gluon fields with nonsinglet flavor quantum numbers
bound to a static color-octet source.
The conserved quantum numbers for the light fields in the presence of the source 
are $J^P_{\rm light}$ and the flavor quantum numbers.
The charge conjugation operator $C_{\rm light}$ 
that changes a $q_1 \bar q_2$ configuration into a 
 $q_2 \bar q_1$ configuration is also a symmetry operator.
Foster and Michael have calculated the adjoint meson spectrum 
using quenched lattice QCD 
with a light valence quark and antiquark  \cite{Foster:1998wu}.  
The adjoint meson energies were calculated for two values
of the common mass of the light valence quark $q$ and antiquark $\bar q$, 
one comparable to the physical mass of the $s$ quark and one larger. 
This allowed for an extrapolation to 
the very small mass of the $u$ and $d$ quarks.  
The adjoint mesons with the lowest energies were found to be
a vector with $J^{PC}_{\rm light}= 1^{--}$ 
and a pseudoscalar with $J^{PC}_{\rm light}= 0^{-+}$.  
Their energies relative to that of the ground-state $1^{+-}$ gluelump
are given in Table~\ref{tab:gl,am}.
For $s \bar s$ adjoint mesons, 
the difference between the energies of the pseudoscalar and vector 
was $50 \pm 70$~MeV, so the vector is favored to be lower in energy.  
The extrapolation of this energy difference to light $q \bar q$ had larger error bars. 
The energy of the $s \bar s$ vector adjoint meson is larger than that of the 
$q \bar q$ vector  adjoint meson by $73 \pm 55$~MeV.
The difference between the energies of the $q \bar q$ vector adjoint meson 
and the ground-state $1^{+-}$ gluelump was $50 \pm 90$~MeV, 
favoring the gluelump to be lower in energy. 
The statistical errors in the energies of the ground-state gluelump, 
the light vector adjoint meson, and the light pseudoscalar adjoint meson
are larger than the energy differences, so the ordering of their energies 
in quenched lattice QCD has not yet been established.

In QCD with two flavors of very light quarks $u$ and $d$,
the lightest adjoint mesons form an isospin triplet with $I=1$ 
and an isospin singlet with $I=0$.
For the isospin singlet and the neutral member of the isospin triplet, 
the $J^{PC}_{\rm light}$ quantum number are $1^{--}$ for the vector
and $0^{-+}$ for the pseudoscalar.
The appropriate quantum numbers for the charged adjoint mesons 
are $I^G(J^P_{\rm light})$, where $G = (-1)^I C_{\rm light}$ 
and $C_{\rm light}$ is the charge conjugation quantum number 
of the neutral member of the multiplet.
The vector adjoint mesons have quantum numbers $0^-(1^-)$
and $1^+(1^-)$.  The pseudoscalar adjoint mesons have 
quantum numbers $0^+(0^-)$ and $1^-(0^-)$.
Calculations of the adjoint meson spectrum using lattice QCD 
with dynamical light quarks are required to determine the ordering 
in energy of the ground-state gluelump 
and the four lowest-energy adjoint mesons with
quantum numbers are $0^-(1^-)$, $1^+(1^-)$,
$0^+(0^-)$, and $1^-(0^-)$.
The energies of these adjoint mesons can be determined 
by a simple minimal-energy prescription if they do not exceed
the energy of the ground-state gluelump by more than 
$2m_\pi$,  $m_\pi$, $2m_\pi$, and $3m_\pi$, respectively.

The existence of a static adjoint meson bound to a local color-octet source
guarantees the existence of corresponding tetraquark potentials
in the small-$r$ region.  Their behavior in this region is that of the 
repulsive color-Coulomb potential for a color-octet $Q \bar Q$ pair, 
analogous to Eq.~(\ref{VGamma:smallR}).  The additive constant
analogous to $E_\Gamma$ can be interpreted as the energy of the adjoint meson.
Given the quantum numbers $J^{P}_{\rm light}$ of an adjoint meson,
we can deduce the B-O potentials for which 
the additive constant $E_\Gamma$ defined by Eq.~(\ref{VGamma:smallR})
is equal to the energy of the  adjoint meson.
A component of the angular momentum vector for an  adjoint meson with spin
$J_{\rm light}$ has $2J_{\rm light} + 1$ integer values ranging
from $-J_{\rm light}$ to $+J_{\rm light}$.  There must therefore be a
B-O potential for each integer value of $\Lambda$ from 0 up to $J_{\rm light}$.
The quantum number $\eta$ for the B-O potentials is the value of 
$(CP)_{\rm light}$ for a $q \bar q$ adjoint meson.
One of the B-O potentials is a $\Sigma$ potential with $\Lambda = 0$.
Its reflection quantum number is
$\epsilon =  (-1)^{J_{\rm light}} P_{\rm light}$.
The B-O potentials whose additive constant as $r \to 0$ equals the energy of the 
vector adjoint meson with $J^P_{\rm light} = 1^-$ are 
$\Pi_g$ and $\Sigma_g^+$.
The B-O potential whose additive constant is equal to the energy of the 
pseudoscalar adjoint meson with $J^P_{\rm light} = 0^-$ is $\Sigma_u^-$.

The behavior of the tetraquark potentials as $r$ increases is not known.
If a $q_1 \bar q_2$ B-O potential can be defined at large $r$, 
the light-field configuration could be a flux tube extending between 
the $Q$ and $\bar Q$ sources to which an excitation
with the flavor quantum numbers $q_1 \bar q_2$ is bound. 
In this case, the B-O potential would increase linearly at large $r$. 
One possibility is that the flavor $q_1$ is localized near the $Q$ source
to form a diquark with color charge $\bar 3$,
and that the flavor $\bar q_2$ is localized near the $\bar Q$ source
to form an antidiquark with color charge $3$.
In this case, the flux tube between the diquark and the antidiquark
would be essentially the same as a flavor-singlet flux tube 
between $\bar Q$ and $Q$ sources.
At large $r$, the B-O potential should approach the energy of a 
relativistic string as in Eq.~(\ref{VGamma:largeR})
for some appropriate excitation number $n_\Gamma$ and
with a different energy offset to account for the energy difference
between the $Q q$ diquark and a $Q$ source.
We will assume that tetraquark potentials can be defined for all $r$,
and that they have the same qualitative behavior 
as the hybrid potentials, with a minimum at a positive value of $r$.

We can use the information from quenched lattice QCD calculations
on the lowest-energy adjoint mesons to infer which tetraquark 
potentials are likely to be the deepest.
The hybrid potentials that are the lowest at small $r$
are also the deepest hybrid potentials. 
We will assume that the tetraquark potentials have the same behavior.
The tetraquark potentials that are the lowest at small $r$ are
$\Pi_g$, $\Sigma_g^+$, and $\Sigma_u^-$.  
They are therefore also likely to be the deepest tetraquark potentials.  
There should be $\Pi_g$, $\Sigma_g^+$, and $\Sigma_u^-$ potentials
for each of the flavor labels $I=0$, $I=1$, $s \bar q$, and $s \bar s$.
We will assume that these tetraquark potentials have the same 
qualitative behavior as the hybrid potentials,  with a minimum at a positive value of $r$.

In Ref.~\cite{Braaten:2013boa}, simple assumptions on the behavior of the 
isospin-1 B-O potentials were used to estimate masses for 
quarkonium tetraquarks. 
The deepest  isospin-1 B-O potentials were assumed to be the same 
as for the flavor-singlet case, namely $\Pi_u$ and $\Sigma_u^-$. 
This assumption was simply a guess, with no motivation from QCD.
The $\Sigma_u^-$ potential coincides with one of the three deepest 
tetraquark potentials inferred from the lowest energy adjoint mesons.

\section{Born-Oppenheimer Energy Levels}
\label{sec:SchrEq}

In this section, we discuss the energy levels of a $Q \bar Q$ pair in
the Born-Oppenheimer (B-O) potentials, which can be identified with $Q \bar Q$ mesons.
We deduce the quantum numbers of the $Q \bar Q$ mesons
and we also derive selection rules for hadronic transitions between them.

\subsection{Angular momenta}
\label{sec:quantum}

When we take into  account the motion and spin of the heavy quark and antiquark,
there are several angular momenta that contribute to the spin vector $\bm{J}$ of the meson.
In addition to the total angular momentum $\bm{J}_{\rm light}$ of the gluon and light-quark fields,
there is the orbital angular momentum $\bm{L}_{Q \bar Q}$ of the $Q \bar Q$ pair
and the spins of the $Q$ and $\bar Q$.  We denote the total spin of the $Q \bar Q$ pair by $\bm{S}$.  
It is convenient  to introduce an angular momentum $\bm{L}$ that is the sum of 
all the angular momenta excluding the spins of the heavy quark and antiquark.
The spin vector of the meson can then be expressed as
\begin{subequations}
\begin{eqnarray}
\bm{J} &=& \bm{L} + \bm{S},
\label{J-LS}
\\
\bm{L} &=&  \bm{L}_{Q \bar Q} + \bm{J}_{\rm light} .
\label{L-LJ}
\end{eqnarray}	
\label{L-LJS}
\end{subequations}
The condition that $\bm{L}_{Q \bar Q}$ is orthogonal to the separation
vector $\bm{r}$  of the  $Q$ and $\bar Q$ can be expressed as 
\begin{equation}
\hat{\bm{r}} \cdot \bm{L} =
 \hat{\bm{r}} \cdot   \bm{J}_{\rm light} = \lambda,
\label{r.L}
\end{equation}	
where $\lambda$ is the quantum number introduced in Section~\ref{sec:lightQN}.
The constraint in Eq.~(\ref{r.L}) puts a lower limit on the quantum number $L$ for
$\bm{L}^2$: $L \ge \Lambda$, where $\Lambda = |\lambda|$. 
		
The centrifugal energy of the $Q \bar Q$ pair
is proportional to the square of their orbital angular momentum:
\begin{equation}
\bm{L}_{Q \bar Q}^2 =  \bm{L}^2
- 2 \bm{L} \cdot \bm{J}_{\rm light}  + \bm{J}_{\rm light}^2.
\label{<L^2>}
\end{equation}	
Imposing the constraint in Eq.~(\ref{r.L}),
this can be expressed as
\begin{equation}
\bm{L}_{Q \bar Q}^2 =  
\bm{L}^2 - 2 \Lambda^2 + \bm{J}_{\rm light}^2 
- \left( L_+ J_{{\rm light},-} + L_- J_{{\rm light},+} \right),
\label{<L^2>-2}
\end{equation}	
where $L_+ $ and $L_-$ are raising and lowering operators for
$\hat{\bm{r}} \cdot \bm{L}$ and $J_{{\rm light},+}$ and $J_{{\rm light},-}$ 
are raising and lowering operators for $\hat{\bm{r}} \cdot   \bm{J}_{\rm light}$.

\subsection{Schroedinger equation}
\label{sec:SchEq}

The B-O approximation consists of two distinct approximations.
The first approximation is an {\it adiabatic approximation},
in which the instantaneous configuration of the gluon and light-quark fields 
is assumed to be a stationary state in the presence of static sources 
at the positions of the $Q$ and $\bar Q$. 
The stationary states can be labelled by the quantum numbers 
$\Gamma = \Lambda_\eta^\epsilon$ introduced in Section~\ref{sec:lightQN}
and by light-quark flavor quantum numbers.
This approximation reduces the problem to a multi-channel nonrelativistic 
Schroedinger equation for the $Q$ and $\bar Q$.
The multi-component wavefunction has a component for every
B-O configuration $\Gamma$ allowed by the symmetries of QCD.
The discrete solutions to the multichannel Schroedinger equation 
correspond to $Q \bar Q$ mesons with definite $J^P$ quantum numbers.
This adiabatic approximation ignores effects that are suppressed by powers of 
$\Lambda_{\rm QCD}/m_Q$ and by powers of $v^2$, 
where $v$ is the typical relative velocity of the $Q \bar Q$ pair,
so it becomes increasingly accurate as the heavy quark mass increases.
The second approximation is a {\it single-channel approximation} in which
all components of the wavefunction are ignored except that for a single 
B-O configuration $\Gamma$.  This approximation breaks down in regions
of $r$ where the B-O potential for $\Gamma$ has avoided crossings with other 
B-O potentials.  It can only be a good approximation if the wavefunction is 
sufficiently small in those regions.

With the combination of the  adiabatic approximation 
and the single-channel approximation, the Schroedinger equation for 
the $Q \bar Q$ pair in the presence of a stationary configuration $\Gamma$ of the 
gluon and light-quark fields  can be expressed as
\begin{equation}
\left[ - \frac{1}{m_Q} \langle \bm{D}^2 \rangle_{\Gamma,\bm{r}} 
+ V_\Gamma(r) \right] \psi(\bm{r})
= E \psi(\bm{r}),
\label{scheq1}
\end{equation}	
where the subscript $\Gamma,\bm{r}$ on the expectation value implies that it is
evaluated in the configuration $\Gamma$ for $Q$ and $\bar Q$ sources 
that are separated by $\bm{r}$.
The covariant derivative $\bm{D}$ has a term with a gluon field that is responsible for
retardation effects.  Since retardation effects are suppressed by powers of $v$,
ignoring these terms is consistent with the adiabatic approximation.
The covariant Laplacian $\bm{D}^2$ can therefore be replaced by an ordinary Laplacian,
which includes a centrifugal term proportional to $\bm{L}_{Q \bar Q}^2$:
\begin{equation}
\left[ - \frac{1}{m_Q} \left( \frac{d \ }{dr} \right)^2 
+ \frac{\langle  \bm{L}_{Q \bar Q}^2 \rangle_{\Gamma,\bm{r}} }{m_Q r^2} 
+ V_\Gamma(r) \right] r \psi(\bm{r})
= E r \psi(\bm{r}).
\label{scheq2}
\end{equation}	
In the expression for $\bm{L}_{Q \bar Q}^2$ in Eq.~(\ref{<L^2>-2}),
the last term is a linear combination of $J_{{\rm light},+}$ and $J_{{\rm light},-}$.
In the multi-channel Schroedinger equation, these terms provide couplings to 
other components of the wavefunction with $\Lambda$ larger by 1 or smaller by 1.
The single-channel approximation eliminates any contribution from these terms.
If the wavefunction is an eigenstate of $\bm{L}^2$
with angular momentum quantum number $L$,
the expectation value of $\bm{L}_{Q \bar Q}^2$ can be expressed as
\begin{equation}
\langle \bm{L}_{Q \bar Q}^2 \rangle_{\Gamma,\bm{r}} =  
L(L+1) - 2 \Lambda^2 + \langle \bm{J}_{\rm light}^2 \rangle_{\Gamma,\bm{r}}.
\label{<L^2>Gamma}
\end{equation}	
Since $\bm{J}_{\rm light}^2$ is a scalar operator,
the function $\langle \bm{J}_{\rm light}^2 \rangle_{\Gamma,\bm{r}}$	
depends on $r$ only. A wavefunction $\psi(\bm{r})$ 
that is a simultaneous eigenstate of $\bm{L}^2$ and $L_z$
with angular momentum quantum numbers $L$ and $m_L$ 
can be expressed in the form $R(r) Y_{Lm_L}(\hat{\bm{r}})$,
where $R(r)$ is a radial wavefunction and $Y_{Lm_L}(\hat{\bm{r}})$
is a spherical harmonic.  The Schroedinger equation in Eq.~(\ref{scheq2}) 
then reduces to the radial Schroedinger equation
\begin{equation}
\left[ - \frac{1}{m_Q} \left( \frac{d \ }{dr} \right)^2 
+ \frac{L(L+1) - 2 \Lambda^2 + \langle \bm{J}_{\rm light}^2 \rangle_{\Gamma,\bm{r}} }
          {m_Q r^2} 
 + V_\Gamma(r) \right] r R(r)
= E r R(r).
\label{scheq3}
\end{equation}	

In the pioneering work on the B-O approximation 
for quarkonium hybrids in Ref.~\cite{Juge:1999ie}, the authors  assumed
without much justification
that $\langle \bm{J}_{\rm light}^2 \rangle_{\Gamma,\bm{r}}$ was
0 for the quarkonium potential $\Sigma_g^+$ and 2 
for the quarkonium hybrid potentials $\Pi_u$ and $\Sigma_u^-$. 
There is a lower bound on the expectation value
in a state with $\hat{\bm{r}} \cdot   \bm{J}_{\rm light} = \lambda$:
$\langle \bm{J}_{\rm light}^2 \rangle \ge \Lambda (\Lambda + 1)$.
The authors assumed that this lower bound is saturated 
in the case of $\Sigma_g^+$, for which $\Lambda = 0$,
and in the case of  $\Pi_u$, for which $ \Lambda = 1$.  
Their assumption that 
$\langle \bm{J}_{\rm light}^2 \rangle _{\Gamma,\bm{r}}= 2$ 
for  $\Pi_u$ and $\Sigma_u^-$ is consistent with a constituent-gluon model
in which the $Q \bar Q$ pair is accompanied by a spin-1 constituent gluon
with $\bm{J}_{\rm light}^2 = 2$.

There is a more compelling motivation for setting 
$\langle \bm{J}_{\rm light}^2 \rangle _{\Gamma,\bm{r}}= 2$ 
for  $\Pi_u$ and $\Sigma_u^-$. 
The centrifugal term in the energy is most important at small $\bm{r}$,
where it provides a centrifugal barrier.  At $\bm{r} = 0$,
the light-field configurations for both $\Pi_u$ and $\Sigma_u^-$
reduce to the ground-state gluelump with quantum numbers $1^{+-}$.
The gluelump is an eigenstate of $\bm{J}_{\rm light}^2$ with eigenvalue 2.
Thus the function $\langle \bm{J}_{\rm light}^2 \rangle _{\Gamma,\bm{r}}$
must be equal to 2 at $\bm{r} = 0$.
In order for $\langle \bm{J}_{\rm light}^2 \rangle_{\Gamma,\bm{r}} \approx 2$ 
to give a good approximation to the solution of the Schroedinger equation, 
it is not necessary for it to be a good approximation to the function at all $\bm{r}$.
It only needs to be a good approximation in the region of small $\bm{r}$
where the centrifugal term is important.  
It should be a good approximation in that region if the light-field configuration
only departs slowly from that of the gluelump as $r$ increases.
The same reasoning applied to a potential $V_\Gamma(r)$
whose additive constant as $r \to 0$ is the energy of a gluelump 
with spin $J_\Gamma$
implies that the appropriate approximation is 
\begin{equation}
\langle \bm{J}_{\rm light}^2 \rangle_{\Gamma,\bm{r}} 
\approx J_\Gamma (J_\Gamma +1).
\label{<Jlight^2>}
\end{equation}	
For the $\Pi_g$ and $\Sigma_g^+{}'$ potentials, whose additive constant 
as $r \to 0$ is the energy of the $1^{--}$ gluelump, the appropriate approximation 
is $\langle \bm{J}_{\rm light}^2 \rangle \approx 2$.
For the $\Delta_g$, $\Pi_g'$, and $\Sigma_g^-{}'$ potentials, 
whose additive constant as $r \to 0$ is the energy of the $2^{--}$ gluelump,
the appropriate approximation is 
$\langle \bm{J}_{\rm light}^2 \rangle \approx 6$.
Similar logic can be applied to the Schroedinger equation for $Q \bar Q$ mesons 
with light-quark flavors.  In this case, $J_\Gamma$ would be the spin 
of the static adjoint meson whose energy determines the additive constant
in the tetraquark potential as $r \to 0$.

The approximation in Eq.~(\ref{<Jlight^2>})
should be a good one if the light-field configuration departs
slowly from the gluelump as $r$ increases.  
Accurate parametrizations of the $\Pi_u$  and $\Sigma_u^-$ potentials
are given in Eqs.~(\ref{VPiu:intR}) and (\ref{VSigmau:intR}).
The absence of linear terms in $r$ implies that the energies of the 
$\Pi_u$  and $\Sigma_u^-$ configurations  remain close to 
the energy of the ground-state gluelump 
until $r$ becomes comparable to $r_0$.  
This is consistent with the assumption that the $\Pi_u$  and $\Sigma_u^-$ 
configurations themselves remain close to the gluelump
until $r$ becomes comparable to $r_0$. 
This suggests that the approximation in Eq.~(\ref{<Jlight^2>})
is probably very good in the short-distance region where the 
centrifugal term in the potential is most important.

When the approximation for 
$\langle \bm{J}_{\rm light}^2 \rangle_{\Gamma,\bm{r}}$ in Eq.~(\ref{<Jlight^2>}) 
is inserted into the radial Schroedinger equation  in Eq.~(\ref{scheq3}),
the function in the numerator of the centrifugal term becomes a number:
\begin{equation}
\left[ - \frac{1}{m_Q} \left( \frac{d \ }{dr} \right)^2 
+ \frac{L(L+1) - 2 \Lambda^2 + J_\Gamma (J_\Gamma +1)}{m_Q r^2} 
+ V_\Gamma(r) \right] r R(r)
= E r R(r).
\label{scheq4}
\end{equation}	
The possible values of the orbital-angular-momentum quantum number 
$L$ are $\Lambda,\Lambda+1,\ldots$.
The radial excitations can be labelled by a principal quantum number 
$n = 1,2,3,\ldots$.

\subsection{Meson quantum numbers}

For each B-O configuration $\Gamma$,
the solution to the radial Schroedinger equation in Eq.~(\ref{scheq4})
gives energy levels $E_{nL}$ and wavefunctions
$R_{nL}(r) Y_{L m_L}(\hat{\bm{r}})$.
The hybrid configurations are labelled by $\Gamma=\Lambda_\eta^\epsilon$.
The energy levels  correspond to configurations of the $Q$ and $\bar Q$ 
and the light fields of the form
\begin{equation}
\big| n L m_L S m_S; \Lambda,\eta, \epsilon \big\rangle
= \int d^3r \, R_{nL}(r) Y_{L m_L}(\hat{\bm{r}})
\big| \Lambda,\eta, \epsilon\,; \bm{r} \big\rangle
\big| S m_S \big\rangle,
\label{ketLS}
\end{equation}	
where $| \Lambda,\eta, \epsilon\,; \bm{r} \rangle$ is the light-field configuration
defined in Eq.~(\ref{Gamma_eta^pm})
and $| S m_S \rangle$ is the spin state of the $Q \bar Q$ pair,
which can be singlet ($S=0$) or triplet ($S=1$).
The state in Eq.~(\ref{ketLS}) is an eigenstate of $P$ and $C$:
\begin{subequations}
\begin{eqnarray}
P &=& \epsilon\,  (-1)^{\Lambda+L+1},
\label{Phybrid}
\\
C &=&  \eta\,  \epsilon\,  (-1)^{\Lambda+L+S}.
\label{Chybrid}
\end{eqnarray}	
\label{PChybrid}%
\end{subequations}
The eigenvalue of $CP$ is the product of $\eta$
for the light-field configuration and $(-1)^{S+1}$ for the spin state of the $Q \bar Q$ pair.
The eigenvalue of the parity operator $P$ is the product of 
$\epsilon (-1)^\Lambda$ for the light-field configuration,
$(-1)^L$ for the spherical harmonic, and  $-1$ 
for the opposite intrinsic parities of the $Q$ and $\bar Q$.

Mesons are states with definite quantum numbers 
for the angular momentum $\bm{J} = \bm{L} + \bm{S}$.
For flavor-singlet $Q \bar Q$ mesons,
the configurations of the $Q$ and $\bar Q$ and light fields with definite 
angular-momentum quantum numbers $J$ and $m_J$ are linear combinations 
of those in Eq.~(\ref{ketLS}) with Clebsch-Gordan coefficients:
\begin{equation}
\big| n L S J m_J; \Lambda,\eta, \epsilon \big\rangle
= \sum_{m_L m_S} \langle L m_L,S m_S | J m_J \rangle
\big| n Lm_L S m_S; \Lambda,\eta, \epsilon \big\rangle.
\label{ketJmJ}
\end{equation}	
In the spin-singlet case ($S=0$), $J$ equals $L$.
In the spin-triplet case ($S=0$), $J$ ranges from $|L-1|$ to $L+1$ in integer steps.
The parity and charge conjugation quantum numbers $P$ and $C$ for the
meson are given in Eqs.~(\ref{PChybrid}).

\begin{table}[t]
\begin{center}
\begin{tabular}{ccc|ccc}
\multicolumn{3}{c}{quarkonia and hybrids} & \multicolumn{3}{c}{$q \bar q$ tetraquarks} \\
~$\Gamma(nL)$~ & $S=0$ & $S=1$ &~$\Gamma(nL)$~ & $S=0$ & $S=1$ \\
\hline
$\Sigma_g^+(1S)$ & $0^{-+}$        & $1^{--}$                         & 
$\Pi_g^-(1P)$         & $1^{+-}$        & $(0,1,2)^{++}$               \\
$\Sigma_g^+(1P)$ & $1^{+-}$         & $(0,1,2)^{++}$              &
$\Pi_g^-(1D)$        & $2^{-+}$          & $(1,2,3)^{--}$                \\
$\Sigma_g^+(1D)$ & $2^{-+}$        & $(1,2,3)^{--}$                         & 
$\Pi_g^-(1F)$         & $3^{+-}$        & $(2,3,4)^{++}$               \\
\hline
$\Pi_u^+(1P)$        & $1^{--}$         & $(0,{\bf 1},2)^{-+}$        & 
$\Pi_g^+(1P)$        & ${\bf 1}^{-+}$ & $({\bf 0},1,2)^{--}$         \\
$\Pi_u^+(1D)$        & $2^{++}$        & $(1,{\bf 2},3)^{+-}$       &
$\Pi_g^+(1D)$        & ${\bf 2}^{+-}$  & $(1,2,3)^{++}$             \\
$\Pi_u^+(1F)$        & $3^{--}$         & $(2,{\bf 3},4)^{-+}$        & 
$\Pi_g^+(1F)$        & ${\bf 3}^{-+}$ & $(2,3,4)^{--}$         \\
\hline
$\Pi_u^-(1P)$         & $1^{++}$       & $({\bf 0},1,{\bf 2})^{+-}$ & 
$\Sigma_g^+(1S)$ & $0^{-+}$        & $1^{--}$                         \\
$\Pi_u^-(1D)$        & $2^{--}$          & $({\bf 1},2,{\bf 3})^{-+}$ &
$\Sigma_g^+(1P)$ & $1^{+-}$         & $(0,1 ,2)^{++}$             \\
$\Pi_u^-(1F)$         & $3^{++}$       & $({\bf 2},3,{\bf 4})^{+-}$ & 
$\Sigma_g^+(1D)$ & $2^{-+}$        & $(1,2,3)^{--}$                 \\
\hline
$\Sigma_u^-(1S)$  & $0^{++}$       & $1^{+-}$                        & 
$\Sigma_u^-(1S)$ & $0^{++}$        & $1^{+-}$                        \\
$\Sigma_u^- (1P)$ & $1^{--}$         & $(0,{\bf 1},2)^{-+}$        &
$\Sigma_u^-(1P)$  & $1^{--}$         & $(0,{\bf 1},2)^{-+}$        \\
$\Sigma_u^-(1D)$  & $2^{++}$       & $(1,{\bf 2},3)^{+-}$        & 
$\Sigma_u^-(1D)$ & $2^{++}$        & $(1,{\bf 2},3)^{+-}$        \\
\end{tabular}
\end{center}
\caption{Spin-symmetry multiplets for the ground state 
and the first two orbital-angular-momentum excitations 
in the quarkonium potential $\Sigma_g^+$,
the two deepest hybrid potentials $\Pi_u$ and $\Sigma_u^-$,
and the $\Pi_g$, $\Sigma_g^+$, and $\Sigma_u^-$ potentials for $q \bar q$ tetraquarks.
A bold $\bm{J}$ indicates that $\bm{J}^{PC}$
is an exotic quantum number that is not possible
if the constituents are only $Q \bar Q$.}
\label{tab:multiplets}
\end{table}

The $Q \bar Q$ mesons are conveniently organized into spin-symmetry multiplets 
consisting of states with the same B-O configuration $\Gamma=\Lambda_\eta^\epsilon$,
radial quantum number $n$,
orbital-angular-momentum quantum number $L$, and flavor.
The states in these multiplets are related by heavy-quark spin symmetry. 
Ordinary quarkonia are energy levels in the flavor-singlet $\Sigma_g^+$ potential.
We set $\Lambda = 0$ and $J_\Gamma=0$ in the Schroedinger equation in Eq.~(\ref{scheq4}).
The possible values of $L$ are $0,1,2,\ldots$ (or equivalently $S,P,D, \ldots$). 
The spin-symmetry multiplets for the ground state $1S$ and the first two 
orbital-angular-momentum excitations $1P$ and $1D$ are
given in Table~\ref{tab:multiplets}.

The lowest-energy quarkonium hybrids are energy levels in the flavor-singlet 
$\Pi_u$ and $\Sigma_u^-$ potentials,
whose short-distance behaviors are determined by the $1^{+-}$ gluelump.
For the deepest hybrid potential $\Pi_u$, we set $\Lambda = 1$ and $J_\Gamma = 1$
in the Schroedinger equation in Eq.~(\ref{scheq4}). 
The possible values of $L$ are $1,2,3,\ldots$ (or equivalently $P,D,F,\ldots$). 
The spin-symmetry multiplets for the ground state $1P$ and the first two 
orbital-angular-momentum excitations $1D$ and $1F$ are given in 
Table~\ref{tab:multiplets} for both the $\Pi_u^+$ and $\Pi_u^-$ configurations.
For the next deepest hybrid potential $\Sigma_u^-$, we set $\Lambda = 0$
and $J_\Gamma = 1$ in the Schroedinger equation in Eq.~(\ref{scheq4}). 
The spin-symmetry multiplets for the ground state $1S$ and the first two 
orbital-angular-momentum excitations $1P$ and $1D$ are
given in Table~\ref{tab:multiplets}.

Tetraquark $Q \bar Q$ mesons are energy levels in potentials 
labelled by quantum numbers $\Gamma=\Lambda_\eta^\epsilon$ 
for the B-O configuration and by flavor quantum numbers.
The flavor labels for distinct B-O potentials are $I=1$, $I=0$, 
$s \bar q$, and $s \bar s$.
The lowest-energy quarkonium tetraquarks are expected to be 
energy levels in the $\Pi_g$, $\Sigma_g^+$, and $\Sigma_u^-$ potentials.
Their multiplets are most easily specified by giving the 
$J^{PC}$ quantum numbers of $q \bar q$ tetraquark mesons.
For the $\Pi_g$ and $\Sigma_g^+$  potentials, $\Lambda$ is 1 and 0, respectively,
and $J_\Gamma = 1$, 
because the short-distance behavior is determined by the $1^{--}$ adjoint meson. 
For the $\Sigma_u^-$ potential,  $\Lambda = 0$ and $J_\Gamma = 0$, 
because the short-distance behavior is determined by the $0^{-+}$ adjoint meson.
The spin-symmetry multiplets for the ground state and the first two 
orbital-angular-momentum excitations of the
$\Pi_g^-$, $\Pi_g^+$, $\Sigma_g^+$, and $\Sigma_u^-$ configurations 
are given in Table~\ref{tab:multiplets}.

 The $J^{PC}$ quantum numbers for $q \bar q$ tetraquarks in Table~\ref{tab:multiplets}
apply to the $I=0$ tetraquark, the $s \bar s$ tetraquark, 
and the neutral member of the $I=1$ isospin triplet.
The isospin triplet has spin-symmetry multiplets whose states have $G$-parity
$G = -C$ and the $J^{P}$ quantum numbers 
of the $q \bar q$ tetraquarks in Table~\ref{tab:multiplets}.
The strange tetraquark mesons containing $u \bar s$ or $d \bar s$
form isospin doublets whose spin-symmetry multiplets 
have the $J^{P}$ quantum numbers 
of the $q \bar q$ tetraquarks in Table~\ref{tab:multiplets}.

\subsection{Selection rules}
\label{sec:selection}

Many of the decay modes of the $XYZ$ mesons listed in 
Tables~\ref{tab:ccneutral}, \ref{tab:cccharged}, and \ref{tab:bbboth}
are hadronic transitions to another $Q \bar Q$ meson.  
Selection rules for the hadronic transitions
provide essential constraints on the quarkonium hybrids 
or quarkonium tetraquarks that can be considered as candidates
for specific $XYZ$ mesons.  The selection rules govern changes in the 
angular momentum quantum numbers $L$, $S$, and $J$ of the $Q \bar Q$ meson
and changes in the quantum numbers $\Lambda$, $\eta$, and $\epsilon$
that specify the light-field configuration.  For simplicity,
we will deduce the selection rules for transitions between neutral $Q \bar Q$ mesons 
with definite $J^{PC}$ quantum numbers.  The corresponding selection rules 
involving charged tetraquark mesons that belong to an isospin triplet
with quantum numbers $1^G(J^P)$
can be inferred from the selection rules involving the neutral member 
of the isospin triplet, whose charge conjugation quantum number is $C = -G$.

There are some selection rules that follow from the exact symmetries of QCD.
These symmetries include rotational symmetry, parity, and charge conjugation.
We take the quantum numbers of the  $Q \bar Q$ mesons before and after
the transition to be $J^{PC}$ and $J^{'P'C'}$.  
We consider a transition via  
the emission of a single hadron $h$ with quantum numbers $J_h^{P_h C_h}$
in a state with orbital-angular-momentum quantum  number $L_h$.  
The conservation of parity and charge conjugation imply the selection rules 
\begin{subequations}
\begin{eqnarray}
P &=& P' P_h (-1)^{L_h} ,
\label{Pmeson}
\\
C &=& C' C_h .
\label{Cmeson}
\end{eqnarray}	
\label{PCmeson}%
\end{subequations}
Angular momentum conservation requires $J$ to be in the range between
$|J' - (J_h+L_h)|$ and $J' + (J_h+L_h)$.

The remaining selection rules for hadronic transitions are only approximate.
There is a {\it spin selection rule} that follows from the approximate 
heavy-quark spin symmetry:
\begin{equation}
S = S',
\label{DeltaS}
\end{equation}	
where $S$ and $S'$ are the total spin quantum numbers for the $Q \bar Q$ pair
before and after the transition.
Transitions between spin-singlet and spin-triplet states 
have rates that are suppressed by the square of
the ratio of a hadronic scale to the heavy quark mass.
Since $m_b$ is about 3 times larger than $m_c$,
this suppression factor is an order of magnitude smaller for 
$b \bar b$ mesons than for $c \bar c$ mesons.

There are also {\it Born-Oppenheimer selection rules} 
that constrain the quantum numbers $\Lambda_\eta^\epsilon$
of the light-field configurations of the $Q \bar Q$ mesons 
involved in the hadronic transition \cite{Braaten:2014ita}.
Since the time scale for evolution of the gluon and light-quark fields
is much faster than that for the motion of the $Q$ and $\bar Q$,
the emission of light hadrons can proceed through an almost instantaneous 
transition of the light-field configuration,
with the positions of the $Q$ and $\bar Q$ remaining essentially fixed.
We consider a transition via  the emission of a single hadron $h$ 
with quantum numbers $J_h^{P_h C_h}$.
The light-field configurations can be labelled by the separation vector 
$\bm{r}$ of the $Q$ and $\bar Q$ and by the quantum numbers introduced in 
Section~\ref{sec:lightQN}:  the eigenvalues $\Lambda$, $\eta$,  and $\epsilon$
of $|\bm{r} \cdot \bm{J}_{\rm light}|$,  $(CP)_{\rm light}$, and $R_{\rm light}$,
respectively.  The flavor-singlet configurations can be denoted by 
kets $| \Lambda, \eta, \epsilon\, ; \bm{r} \rangle$.
For $\Lambda \ge 1$, these kets are linear combination of eigenstates of 
$\bm{r} \cdot \bm{J}_{\rm light}$ with eigenvalues $\lambda=\pm \Lambda$.
A hadronic transition between flavor-singlet light-field configurations 
in which the hadron $h$ is emitted with momentum $\bm{q}$ can be expressed as
\begin{equation}
\big| \Lambda, \eta, \epsilon\, ; \bm{r} \big\rangle \longrightarrow
\big| \Lambda', \eta' , \epsilon\,; \bm{r} \big\rangle 
\big| h(\bm{q}) \big\rangle .
\label{hadtrans}
\end{equation}	
The conservation of the component of the total angular momentum 
$\bm{J}_{\rm light}$ of the light
fields along the $Q \bar Q$ axis can be expressed as
\begin{equation}
\lambda = \lambda'
+ \hat{\bm{r}} \cdot (\bm{J}_h + \bm{L}_h ),
\label{Jlightcons}
\end{equation}	
where $\bm{J}_h$ and $\bm{L}_h$ are the spin vector  and
orbital-angular-momentum vector 
of the light hadron $h$.  If $h$ is emitted with 
orbital-angular-momentum quantum number $L_h$,
the constraint in Eq.~(\ref{Jlightcons}) implies the selection rule
\begin{equation}
|\lambda - \lambda'| \le J_h  + L_h.
\label{lambdaselect}
\end{equation}	
The quantum numbers $\eta$ and $\eta'$ in Eqs.~(\ref{hadtrans})
are the eigenvalues of $(CP)_{\rm light}$  for the light-field configurations.
Conservation of $(CP)_{\rm light}$  implies the selection rule
\begin{equation}
\eta = \eta' \cdot C_h P_h (-1)^{L_h}.
\label{CPselect}
\end{equation}	
In the special case $\lambda = \lambda'=0$, there is an additional constraint
from invariance under reflection through a plane containing the $Q \bar Q$ axis.
The initial and final light-field configurations 
$| 0, \eta, \epsilon\, ; \bm{r} \rangle$ and 
$| 0, \eta', \epsilon'\, ; \bm{r} \rangle$ are eigenstates of the reflection operator
$R_{\rm light}$ with eigenvalues $\epsilon $ and $\epsilon'$, respectively.
The effect of the reflection on the emitted hadron can be deduced by
expressing $R_{\rm light}$ as the product of the parity operator $P_{\rm light}$
and a rotation by angle $\pi$ around the axis of the reflection plane.
Such a rotation changes the phase by 
$\exp(i \pi \hat{\bm{r}} \cdot (\bm{J}_h + \bm{L}_h ))$,
which equals 1 by the constraint in Eq.~(\ref{Jlightcons}).
The additional constraint imposed by the 
reflection symmetry is therefore
\begin{equation}
\epsilon = \epsilon' \cdot P_h (-1)^{L_h}
\qquad (\lambda = \lambda'=0).
\label{Pselect}
\end{equation}	

The hadronic transitions are also governed by flavor selection rules
associated with conservation of net light-quark flavors.	  
We will only consider hadronic transitions in which the final 
$Q \bar Q$ meson is a quarkonium.  Since it is a 
flavor singlet, the flavor selection rules are trivial.

\subsection{Candidates for $\bm{XYZ}$ mesons}
\label{sec:candidates}

Many of the hadronic transitions of the $XYZ$ mesons listed in Tables~\ref{tab:ccneutral}, 
\ref{tab:cccharged}, and \ref{tab:bbboth} are to quarkonium states.
For $c \bar c$ mesons, the charmonium states are the spin-triplet $1^{--}$
states $J/\psi$ and $\psi(2S)$, the spin-triplet $1^{++}$ state $\chi_{c1}(1P)$,
and the spin-singlet $1^{+-}$ state $h_{c1}(1P)$.
For $b \bar b$ mesons, the bottomonium states are the spin-triplet $1^{--}$
states $\Upsilon(nS)$ and the spin-singlet $1^{+-}$ states $h_{c1}(nP)$.

When applied to the $c \bar c$ mesons listed in Tables~\ref{tab:ccneutral} 
and \ref{tab:cccharged}, the spin selection rule implies that
the only plausible candidates for $XYZ$ mesons 
with transitions to $J/\psi$, $\psi(2S)$, or $\chi_{cJ}(1P)$
are spin-triplet members of charmonium hybrid 
or charmonium tetraquark multiplets.
The only plausible candidates for $XYZ$ mesons with transitions to $h_c(1P)$
are spin-singlet members of charmonium hybrid 
or charmonium tetraquark multiplets.
The spin selection rule puts strong constraints on the interpretations 
of the $XYZ$ mesons in Table~\ref{tab:ccneutral} with quantum numbers $1^{--}$.
In the quarkonium hybrid multiplets listed in Table~\ref{tab:multiplets},
the only $1^{--}$ states are the spin-singlet members 
of the $\Pi_u^+(1P)$ and $\Sigma_u^-(1P)$ multiplets.
In the quarkonium tetraquark multiplets listed in Table~\ref{tab:multiplets},
there is a spin-singlet $1^{--}$ state in the $\Sigma_u^-(1P)$ multiplet
and there are spin-triplet $1^{--}$ states in the 
$\Pi_g^-(1D)$, $\Pi_g^+(1P)$,  $\Sigma_g^+(1S)$, and $\Sigma_g^+(1D)$ multiplets.
The $1^{--}$ $c \bar c$ meson $Y(4220)$ in Table~\ref{tab:ccneutral},
which decays into $h_c(1P)\, \pi^+ \pi^-$ \cite{Yuan:2013ffw}, 
must be a spin-singlet.
If we assume that the $Y(4220)$ is the ground state of a B-O potential,
it can only be identified with the $1^{--}$ state 
in the $\Pi_u^+(1P)$ energy level of the charmonium hybrid.
The $1^{--}$ $c \bar c$ meson $Y(4260)$ in Table~\ref{tab:ccneutral}, 
which decays into $J/\psi\, \pi^+\pi^-$  \cite{Aubert:2005rm}, 
must be a spin-triplet.
If we assume that the $Y(4260)$ is the ground state in a B-O potential,
it can be identified with the $1^{--}$ state in either the $\Pi_g^+(1P)$ 
or $\Sigma_g^+(1S)$ energy level of the isospin-0 charmonium tetraquark.

When applied to the $b \bar b$ mesons listed in Table \ref{tab:bbboth},
the spin selection rule presents a puzzle.
The $Z_b^+(10610)$ and $Z_b^+(10650)$ have hadronic transitions
to both the spin-triplet bottomonium states $\Upsilon(nS)$
and  the spin-singlet bottomonium states $h_b(nS)$ \cite{Belle:2011aa}.  
Thus their decays violate the spin selection rule.
This can be explained by the $Z_b^+(10610)$ having a large $B^* \bar B$ 
molecular component and the $Z_b^+(10650)$ having a large $B^* \bar B^*$ 
molecular component \cite{Bondar:2011ev,Cleven:2011gp,Mehen:2011yh}.
Within the Born-Oppenheimer approach, the large molecular components 
would arise from energy levels in B-O potentials that are fortuitously
close to the $B^* \bar B$ and $B^* \bar B^*$ thresholds,
which results in a breakdown of the single-channel approximation.

Several of the hadronic transitions for the neutral $c \bar c$ mesons listed in 
Table~\ref{tab:ccneutral} are the emission of a single vector meson 
$\omega$ or $\phi$ with   $J_h^{P_h C_h} = 1^{--}$.
Since the kinetic energy of the vector meson is small compared to its mass,
we assume it is emitted in an $S$-wave state.
The B-O selection rules in Eqs.~(\ref{lambdaselect}), (\ref{CPselect}),
and (\ref{Pselect}) reduce to
$|\lambda - \lambda'| \le 1$, $\eta = \eta'$, and 
also $\epsilon = - \epsilon'$ if $\lambda = \lambda'=0$.
If the final-state configuration is $\Sigma_g^+$ corresponding to a quarkonium,
the selection rules reduce further to $\Lambda \le 1$, $\eta = +1$, and 
also $\epsilon =  -1$ if $\Lambda=0$.  They imply that the only 
possible initial-state configurations are $\Pi_g^-$, $\Pi_g^+$, and $\Sigma_g^-$.
The quarkonium hybrid configurations with the deepest potentials are
$\Pi_u^+$, $\Pi_u^-$, and $\Sigma_u^-$.  None of these can 
make a transition to quarkonium through the $S$-wave emission of 
a vector meson.  
The quarkonium tetraquark configurations with the deepest potentials are
presumably $\Pi_g^-$, $\Pi_g^+$,  $\Sigma_g^+$, and $\Sigma_u^-$.  
Of these, the only ones that can make a transition to quarkonium 
through the $S$-wave emission of a vector meson are $\Pi_g^-$ and $\Pi_g^+$.
We first consider the $X(3915)$, which decays into $J/\psi\, \omega$
and has quantum numbers  $0^{++}$  \cite{Lees:2012xs}.
In the $\Pi_g^-$ and $\Pi_g^+$ tetraquark multiplets listed in Table~\ref{tab:multiplets}, 
the only $0^{++}$ state is a spin-triplet member of 
$\Pi_g^-(1P)$.  We therefore identify $X(3915)$ with the $0^{++}$ member
of the $\Pi_g^-(1P)$ multiplet of isospin-0 charmonium tetraquarks.
We next consider the $Y(4140)$, $Y(4274)$, and $X(4350)$,
which have $C=+$ and decay into $J/\psi\, \phi$.  The $\phi$ in the final state 
suggests that the meson is an $s \bar s$ tetraquark.
In the $\Pi_g^-$ and $\Pi_g^+$ tetraquark multiplets listed in Table~\ref{tab:multiplets}, 
there are spin-triplet $C=+$ states in the multiplets
$\Pi_g^-(1P)$, $\Pi_g^-(1F)$, and $\Pi_g^+(1D)$.
If we assume that the lowest of these three states, $Y(4140)$, 
is in the ground state of a B-O potential, it must be the 
$0^{++}$, $1^{++}$, or $2^{++}$ member
of the $\Pi_g^-(1P)$ multiplet of $s \bar s$ charmonium tetraquarks.
The energy difference of about 230~MeV between the $Y(4140)$
and the $X(3915)$ is approximately twice the difference between the
constituent masses of an $s$ quark and a lighter quark.
It is therefore compatible with the identifications 
of $Y(4140)$ and $X(3915)$ as states in the $\Pi_g^-(1P)$ 
multiplets of $s \bar s$ and isospin-0 charmonium tetraquarks, respectively.

The hadronic transitions for the charged $c \bar c$ mesons listed in 
Table~\ref{tab:cccharged} and for the charged $b \bar b$ mesons listed in 
Table~\ref{tab:bbboth} are the emission of a single $\pi^+$ with   $J_h^{P_h C_h} = 0^{-+}$.
The Goldstone nature of the pion requires that it be emitted in a $P$-wave state.
The B-O selection rules in Eqs.~(\ref{lambdaselect}), (\ref{CPselect}), and (\ref{Pselect}) 
reduce to $|\lambda - \lambda'| \le 1$, $\eta = \eta'$, and 
also $\epsilon = \epsilon'$ if $\lambda = \lambda'=0$.
If the final-state configuration is $\Sigma_g^+$ corresponding to a quarkonium,
the selection rules reduce further to $\Lambda \le 1$, $\eta = +1$, and 
also $\epsilon =  +1$ if $\Lambda=0$.  They imply that the only 
possible initial-state configurations are $\Pi_g^-$, $\Pi_g^+$, and $\Sigma_g^+$.
Isospin symmetry provides the additional selection rule that the initial configuration 
must have isospin 1.  
The quarkonium tetraquark configurations with the deepest potentials are
presumably $\Pi_g^-$, $\Pi_g^+$,  $\Sigma_g^+$, and $\Sigma_u^-$.  
The only ones that can make a transition to quarkonium 
through the $P$-wave emission of a pion are $\Pi_g^-$, $\Pi_g^+$, 
and $\Sigma_g^+$.
The $Z_c^+(3900)$ decays into $J/\psi\, \pi^+$  \cite{Ablikim:2013mio}.
Its neutral isospin partner $Z_c^0(3900)$ has $C=-$.
The $\Pi_g^-$, $\Pi_g^+$, and $\Sigma_g^+$ tetraquark multiplets 
listed in Table~\ref{tab:multiplets} include many spin-triplet $C=-$ states.  
If we assume the $Z_c^0(3900)$ is in the ground state
of a B-O potential, it must  be the $0^{--}$, $1^{--}$, or $2^{--}$ state in
the $\Pi_g^+(1P)$ multiplet or the $1^{--}$ state in the $\Sigma_g^+(1S)$ 
multiplet of isospin-1 charmonium tetraquarks.
The $Z_c^+(4020)$ decays into $h_c(1P)\, \pi^+$ \cite{Ablikim:2013wzq}.
Its neutral isospin partner $Z_c^0(4020)$ has $C=-$.
In the $\Pi_g^-$, $\Pi_g^+$, and $\Sigma_g^+$ tetraquark multiplets 
listed in Table~\ref{tab:multiplets}, the spin-singlet $C=-$ states 
are in the $\Pi_g^-(1P)$, $\Pi_g^-(1F)$, $\Pi_g^+(1D)$, 
and $\Sigma_g^+(1P)$ multiplets.  If we assume the $Z_c^0(4020)$ is in the ground state
of a B-O potential, it must be the $1^{+-}$ state in the $\Pi_g^-(1P)$ 
multiplet of isospin-1 charmonium tetraquarks.
The $Z_1^+(4050)$ and $Z_2^+(4250)$ 
decay into $\chi_{c1}(1P)\, \pi^+$ \cite{Mizuk:2008me}.  
Their neutral isospin partners $Z_1^0(4050)$ and $Z_2^0(4250)$ have $C=+$.  
The $\Pi_g^-$, $\Pi_g^+$, and $\Sigma_g^+$ tetraquark multiplets 
listed in Table~\ref{tab:multiplets} include many spin-triplet $C=+$ states.  
If we assume the $Z_1^0(4050)$ is in the ground state
of a B-O potential, it must be the $0^{++}$, $1^{++}$, or $2^{++}$ state
in the $\Pi_g^-(1P)$ multiplet of isospin-1 charmonium tetraquarks.
The small difference between the  masses of $Z_c^+(4020)$ and $Z_1^+(4050)$
is compatible with their being different states in the 
$\Pi_g^-(1P)$ multiplet of isospin-1 charmonium tetraquarks.

Finally, we consider the implications of the B-O selection rules in Eqs.~(\ref{lambdaselect}), 
(\ref{CPselect}), and (\ref{Pselect}) for the only neutral $b \bar b$ $XYZ$ meson 
listed in Table~\ref{tab:bbboth}. 
The $Y(10890)$ has quantum numbers $1^{--}$, 
and it has been observed in the decay channel $\Upsilon(nS)\, \pi^+\pi^-$\cite{Adachi:2008pu}.
If this state is a ground-state energy level of a B-O potential, the only
bottomonium hybrid option is the spin-singlet $1^{--}$ state in the 
$\Pi_u^+(1P)$ multiplet.  The bottomonium tetraquark options are the 
spin-triplet $1^{--}$ state in either the $\Pi_g^+(1P)$ multiplet
or the $\Sigma_g^+(1S)$ multiplet.  The decays of  $Y(10890)$ into 
$\Upsilon(nS)\, \pi^+\pi^-$ favors one of the spin-triplet options.
However these decays may have large contributions from direct decays into 
$Z_b(10610)\, \pi$ and $Z_b(10650)\, \pi$, followed by the subsequent decay
of the $Z_b$ meson into $\Upsilon(nS)\, \pi$.
The $Z_b(10610)$ and $Z_b(10650)$ also decay into $h_b(nP)\, \pi$.
Hadronic transitions of the $Z_b$ mesons to both the spin-triplet bottomonium states $\Upsilon(nS)$
and the spin-singlet bottomonium states $h_b(nP)$ violate the spin selection rule.
In order to determine the implications of the B-O selection rules for the $Y(10890)$,
we would first need to determine whether its direct decays obey the spin selection rule.

\section{Lattice Gauge Theory}
\label{sec:lattice}

In this section, we describe the existing results on the spectra
of hybrid and tetraquark $c \bar c$ and $b \bar b$ mesons 
that have been calculated using lattice gauge theory.
They provide some information about the pattern of deviations from the 
Born-Oppenheimer approximation.

\subsection{Charmonium hybrids from lattice QCD}
\label{sec:latticeQCD}

The spectrum of charmonium hybrids
can be calculated directly using lattice QCD.  
Exploratory calculations of the $c \bar c$ meson spectrum 
above the open-charm threshold 
were carried out by Dudek, Edwards, Mathur, and Richards \cite{Dudek:2007wv}
and extended by the Hadron Spectrum Collaboration \cite{Liu:2012ze}.  
The most recent published calculations
used an anisotropic lattice with $24^3 \times 128$ sites
and a spatial lattice spacing of about 0.12~fm.
Their gauge field configurations were generated using dynamical 
$u$, $d$, and $s$ quarks, with the $s$ quark having its physical 
mass and the $u$ and $d$ quarks unphysically heavy, 
corresponding to a pion mass of about 400~MeV.
On a cubic lattice, there are 20 channels analogous to the 
$J^{PC}$ quantum numbers in the continuum.
For each of the 20 lattice $J^{PC}$ channels,
the Hadron Spectrum Collaboration calculated the $c \bar c$ meson spectrum
from the Euclidean time dependence of the cross-correlators
for a set of operators whose number ranged from 4 to 26,
depending on the channel.  The operators included ``hybrid'' 
operators constructed out of the $c$ quark field 
and the gluon field strength and ``charmonium''  
operators constructed out of the $c$ quark field and 
other combinations of covariant derivatives.

The Hadron Spectrum Collaboration identified 46 states 
in the $c \bar c$ meson spectrum 
with high statistical precision \cite{Liu:2012ze}. 
These states had spins $J$  as high as 4 and masses as high as 4.6~GeV.  
The states that are more strongly excited by hybrid operators are 
plausible candidates for charmonium hybrids.
All the charmonium hybrid candidates in the calculations of Ref.~\cite{Liu:2012ze}
can be organized into four complete heavy-quark spin-symmetry multiplets:  
\begin{subequations}
\begin{eqnarray}
H_1 &=& \{ 1^{--}, (0,\bm{1},2)^{-+} \},
\label{H1}
\\
H_2 &=& \{ 1^{++}, (\bm{0},1,\bm{2})^{+-} \},
\label{H2}
\\
H_3 &=& \{ 0^{++}, 1^{+-} \},
\label{H3}
\\
H_4 &=& \{ 2^{++}, (1,\bm{2},3)^{+-} \}.
\label{H4}
\end{eqnarray}	
\label{H1234}%
\end{subequations}
A bold $\bm{J}$ indicates that $\bm{J}^{PC}$
is an exotic quantum number that is not possible
if the constituents are only $Q \bar Q$.
If we consider only the central values of the energies of the states, 
the ordering in energies of the multiplets from lowest to highest are
$H_1$, $H_2$, $H_3$, and $H_4$.  
If we take into account the statistical errors in the energies,
there may be small overlaps between some of the multiplets.
The calculations in Ref.~\cite{Liu:2012ze} are not definitive,
because they were not extrapolated to zero lattice spacing 
or to the physical values of the $u$ and $d$ quark masses.  
However since light quarks are not expected to be important
as constituents in charmonium or in charmonium hybrids, 
the results of Ref.~\cite{Liu:2012ze} provide plausible estimates
for the masses of the charmonium hybrids.  
The energy splittings between charmonium hybrid states 
may be less sensitive to the effects of light quarks than their masses.

\begin{table}[t]
\begin{center}
\begin{tabular}{cccc|cc}
 &  &  &  &  & ~$c \bar c$ hybrids~  \\
~multiplet~ & ~$\Gamma$~ & ~$nL$~ & ~$\langle$energy$\rangle_{\rm spin}$~ & 
                      ~$J^{PC}$~ & ~lattice QCD~  \\
\hline
$H_1$ & ~$\Pi_u^+$~ &   ~$1P$~ & ~$4212 \pm 22$~ & 
                         $1^{--}$ & ~$\bm{(4216 \pm 7)}$~ \\
 & & & &           $0^{-+}$    & ~$4126 \pm 20$~ \\
 & & & & ~$\bm{1^{-+}}$~ & ~$4148 \pm 22$~ \\
 & & & &          $2^{-+}$     & ~$4265 \pm 23$~ \\
\hline
$H_2$ & ~$\Pi_u^-$~        & ~$1P$~ & ~$4314 \pm 32$~ & 
                        $1^{++}$    & ~$4330 \pm 21$~ \\
 & & & & ~$\bm{0^{+-}}$~ & ~$4317 \pm 18$~ \\
 & & & &          $1^{+-}$     & ~$4275 \pm 41$~ \\
 & & & & ~$\bm{2^{+-}}$~ & ~$4326 \pm 43$~ \\
\hline
$H_3$ & ~$\Sigma_u^-$~ & ~$1S$~ & ~$4407 \pm 22$~ & 
                        $0^{++}$     & ~$4403 \pm 34$~ \\
 & & & &          $1^{+-}$     & ~$4408 \pm 25$~ \\
\hline
$H_4$ & ~$\Pi_u^+$~ & ~$1D$~ & ~$4448 \pm 19$~ & 
                        $2^{++}$    & ~$4423 \pm 26$~ \\
 & & & &          $1^{+-}$     & ~$4428 \pm 42$~ \\
 & & & & ~$\bm{2^{+-}}$~ & ~$4443 \pm 24$~ \\
 & & & &          $3^{+-}$     & ~$4479 \pm 20$~ \\
\hline
\end{tabular}
\end{center}
\caption{Charmonium hybrid energies (in MeV) 
predicted using the splittings between states 
calculated using lattice QCD in Ref.~\cite{Liu:2012ze}.
The experimental input in parentheses is the measured mass of the $Y(4220)$,
which is identified as the $1^{--}$ hybrid in the ground-state $H_1$ multiplet.
The column labelled $\langle$energy$\rangle_{\rm spin}$
gives the spin-averaged energies of the multiplets. 
The error bars take into account the statistical errors in the lattice calculations.
They do not account for systematic errors associated with extrapolations 
to zero lattice spacing and to the physical $u$ and $d$ quark masses.}
\label{tab:chybrids}
\end{table}

In Ref.~\cite{Braaten:2013boa}, the $Y(4260)$ 
was identified as the lowest $1^{--}$ charmonium hybrid.
The masses of other charmonium hybrids were then 
estimated by using the results of Ref.~\cite{Liu:2012ze} 
for the splittings between $c \bar c$ mesons. 
This identification was motivated 
primarily by the very small cross section for 
producing the $Y(4260)$ in $e^+ e^-$ annihilation, despite
it having the appropriate quantum numbers $1^{--}$.
The small cross section can be explained by the small 
wavefunction for $c \bar c$ at the origin that is characteristic 
of a quarkonium hybrid.  The only $1^{--}$ state 
among the charmonium hybrid multiplets 
in Eqs.~(\ref{H1234}) is the spin-singlet member of $H_1$.  
But an important decay mode 
of the $Y(4260)$ is the discovery channel $J/\psi\, \pi^+ \pi^-$,
which is a spin-triplet decay channel.
The identification of the $Y(4260)$ with this state 
is therefore disfavored by the spin selection rule, which
requires a spin-singlet charmonium hybrid to decay
preferentially into spin-singlet charmonium states.  

The BESIII Collaboration has recently observed the $Y(4220)$,
which has quantum numbers $1^{--}$ and decays into
$h_c\, \pi^+ \pi^-$, which is a spin-singlet decay channel.
The  $Y(4220)$ can plausibly be identified with the 
spin-singlet member of the $H_1$ multiplet.
We therefore identify the $Y(4220)$ as the $1^{--}$ member of the 
ground-state charmonium hybrid multiplet.
The masses of other charmonium hybrids are then 
estimated by using the results of Ref.~\cite{Liu:2012ze} 
for the splittings between $c \bar c$ mesons. 
The results are shown in Table~\ref{tab:chybrids}. 
The errors are statistical uncertainties only.
They do not include the systematic errors associated 
with the extrapolation to zero lattice spacing 
or to the small physical masses of the $u$ and $d$ quarks.

The Born-Oppenheimer interpretations of the multiplets 
$H_1$, $H_2$, $H_3$, and $H_4$ in Eqs.~(\ref{H1234}) 
are the $\Pi^+_u(1P)$, $\Pi^-_u(1P)$, $\Sigma_u^-(1S)$, 
and $\Pi_u^+(1D)$ energy levels, respectively.
The results in Table~\ref{tab:chybrids} give us some idea of the size of 
corrections to the B-O approximation.
In the leading B-O approximation, the
$\Pi_u$ potential is the same for $\epsilon = + 1$ and $-1$.
However the spin average for the $\Pi^-_u(1P)$ multiplet 
is about 100~MeV higher than the spin average for the 
$\Pi^+_u(1P)$ multiplet.
In the leading B-O approximation, 
the states in each spin-symmetry multiplet are degenerate.
However the range of energies within the $H_1$ multiplet is about 
140~MeV and the ranges of energies within the 
$H_2$ and $H_4$ multiplets are about 60~MeV.
The $\Pi^+_u(1P)$ and $\Pi_u^+(1D)$ energy levels 
are the ground state and the first orbital-angular-momentum excitation 
of the $\Pi^+_u$ configuration.  The splitting of 240~MeV
between their spin averages can be used as an estimate 
for orbital-angular-momentum splittings.

An alternative interpretation of the lowest charmonium hybrid spin-symmetry multiplets 
can be obtained by interpreting the $1^{+-}$ ground-state gluelump
as a constituent gluon with no orbital angular momentum
bound to a color-octet heavy quark-antiquark pair.
If the $Q \bar Q$ pair is in an $S$-wave state, 
its spin-symmetry multiplet is $\{ 0^{-+}, 1^{--} \}$.
The spin-symmetry multiplet of the hybrid meson is then $1^{+-} \otimes \{ 0^{-+}, 1^{--} \}$,
which is equivalent to $H_1 = \{ 1^{--}, (0,\bm{1},2)^{-+} \}$. 
Thus the ground-state gluelump bound to an $S$-wave $Q \bar Q$ pair
gives rise to the same spin-symmetry multiplet as the ground-state $Q \bar Q$ hybrid.
If the $Q \bar Q$ pair is in a $P$-wave state, 
its spin-symmetry multiplet is $\{ 1^{+-}, (0,1,2)^{++} \}$.
The spin-symmetry supermultiplet of the hybrid mesons is 
then $1^{+-} \otimes  \{ 1^{+-}, (0,1,2)^{++} \}$.
The spin-singlet states are $(0,1,2)^{++}$ and the
spin-triplet states are $(\bm{0},1,1,1,\bm{2},\bm{2},3)^{+-}$. 
They account for all the states in $H_2$, $H_3$, and $H_4$.
Dudek has argued that the first excited energy levels of a light-quark hybrid 
form such a supermultiplet \cite{Dudek:2011bn}. 
For the charmonium hybrid energy levels in Table~\ref{tab:chybrids},
the range of energies for the complete supermultiplet consisting of 
$H_2$, $H_3$, and $H_4$ is about 200~MeV, while the ranges of energies
within the individual multiplets $H_2$, $H_3$, and $H_4$ are about 55, 5, 
and 56~MeV, respectively.
This suggest that the Bonn-Oppenheimer interpretation provides a more
useful first approximation than the constituent gluelump interpretation.

\subsection{Bottomonium hybrids from lattice NRQCD}
\label{sec:latticeNRQCD}

The mass of the bottom quark is too large for lattice QCD
with a conventional isotropic lattice to be applied directly to $b \bar b$ mesons
with currently available computational resources.
One alternative is to use an anisotropic lattice in which the lattice spacing
is much finer in the Euclidean time direction than in the three spacial directions.
Another  alternative is to use a lattice discretization of an effective field theory 
called {\it NonRelativistic QCD} (NRQCD)
in which the $b$ quark is treated nonrelativistically \cite{Lepage:1992tx}.
This method can also be applied to $c \bar c$ mesons,
although the errors associated with the nonrelativistic approximation are 
larger. 

Quenched lattice NRQCD was used by Juge, Kuti, and Morningstar 
to calculate the masses of some of the states in the $b \bar b$ meson 
spectrum \cite{Juge:1999ie}.  
They used a lattice NRQCD action that included only the leading 
terms in the velocity expansion, which gives no spin splittings
within spin-symmetry multiplets.
They used a lattice with $15^3 \times 45$ sites 
and a spatial lattice spacing of about 0.12~fm. 
For each of 5 lattice $J^{PC}$ channels,
they calculated the $b \bar b$ meson spectrum
from the Euclidean time dependence of the cross-correlators
for either 1 or 4 operators.
The states that are more strongly excited by hybrid operators are 
plausible candidates for bottomonium hybrids.
They identified four candidate bottomonium hybrid states.
The three lowest-energy states were the $1^{--}$, $1^{++}$, and $0^{++}$
members of the lowest three spin-symmetry multiplets $H_1$, $H_2$, and $H_3$
defined in Eq.~(\ref{H1234}).
The fourth and highest-energy state, which has quantum numbers $1^{--}$,
can be interpreted as a radial excitation of the ground-state quarkonium hybrid.
The corresponding multiplet is labelled $H_1'$.
Candidates for such a multiplet were not observed in 
the lattice QCD calculations of charmonium hybrids in Ref.~\cite{Liu:2012ze}.
The reason for this could be that the operators used to excite
charmonium hybrids in Ref.~\cite{Liu:2012ze} did not couple sufficiently strongly 
to radial excitations.
The difference of about 440~MeV between the masses of the $1^{--}$
states in the $H_1'$ and $H_1$ multiplets can be used as an estimate 
for splittings between radial excitations.

Liao and Manke have calculated the $b \bar b$ meson spectrum using 
quenched lattice QCD on an anisotropic lattice \cite{Liao:2001yh}. 
They used a lattice with $16^3 \times 128$ sites, 
with a spacial lattice spacing of 0.05~fm 
and a much finer lattice spacing in the Euclidean time direction.
They determined the masses for three $b \bar b$ hybrid mesons
with exotic quantum numbers  from the Euclidean time dependence 
of the correlators
of appropriate operators. The $\bm{1^{-+}}$ state is a member 
of the $H_1$ multiplet,
while the $\bm{0^{+-}}$ and $\bm{2^{+-}}$ states are members 
of the $H_2$ multiplet.

\begin{table}[t]
\begin{center}
\begin{tabular}{ccc|ccc}
 &  &  &  &  ~$b \bar b$ hybrids~ &  ~$b \bar b$ hybrids~ \\
~multiplet~ & ~$\Gamma$~ & ~$nL$~ & 
~$J^{PC}$~ &  ~lattice NRQCD~ &  ~lattice QCD~ \\
\hline
$H_1$ & ~$\Pi_u^+$~ & ~$1P$~ & $1^{--}$           & ~(10559)~ &  \\
           &                       &               & $\bm{1^{-+}}$ &                   & ~(10559)~   \\
\hline
$H_2$ & ~$\Pi_u^-$~  & ~$1P$~ & $1^{++}$         & ~$10597\pm 65$~ & \\
           &                       &               & $\bm{0^{+-}}$ &                            & ~$10159\pm 362$~  \\
           &                       &               & $\bm{2^{+-}}$ &                            & ~$11323\pm 257$~  \\
\hline
$H_3$ & ~$\Sigma_u^-$~ & ~$1S$~ & $0^{++}$   & $10892 \pm 36$  & \\
\hline
$H_1'$ & ~$\Pi_u^+$~       & ~$2P$~ & $1^{--}$  & $10977 \pm 41$    & \\
\hline
\end{tabular}
\end{center}
\caption{Bottomonium hybrid energies (in MeV) 
predicted using the splittings between states 
calculated using quenched lattice NRQCD in Ref.~\cite{Juge:1999ie}
and using quenched lattice QCD in Ref.~\cite{Liao:2001yh}.
The inputs in parentheses for the $\Pi_u^+(1P)$ energy levels
have been chosen arbitrarily to be the $B \bar B$ threshold.
The error bars take into account the statistical errors in the lattice calculations
and the uncertainties from setting the heavy-quark mass.
They do not account for systematic errors associated with extrapolations 
to zero lattice spacing and from the omission of light quark loops.}
\label{tab:bhybrids}
\end{table}

Since the lattice calculations in Refs.~\cite{Juge:1999ie} and  \cite{Liao:2001yh} 
do not include the effects of light-quark loops, 
any quantitative predictions should be treated with caution.
Nevertheless, we proceed to use the results to estimate the energy levels for 
bottomonium hybrids.
In the lattice NRQCD calculations of Ref.~\cite{Juge:1999ie}, 
the lattice energy scale cancels out in
the ratios of the energy splittings of the bottomonium hybrids 
to the $1P-1S$ splitting of bottomonium.  
We therefore determine the energy splittings
by multiplying the ratios by the observed $1P-1S$ splitting of bottomonium. 
To determine the absolute energies of the bottomonium hybrids
in the lattice NRQCD calculation,
we need an experimental input to determine the energy offset.
We arbitrarily choose the mass of the $1^{--}$ member of the 
ground-state $H_1$ multiplet to be the $B \bar B$ threshold,
which is 10559~MeV.
The splittings from Ref.~\cite{Juge:1999ie} are then used to estimate
the masses of other bottomonium hybrids, which are  
given in Table~\ref{tab:bhybrids}.
To determine the absolute energies of the bottomonium hybrids
in the lattice QCD calculation,
we arbitrarily choose the mass of the $\bm{1^{-+}}$ member of the 
ground-state $H_1$ multiplet to be the $B \bar B$ threshold.
The splittings from Ref.~\cite{Liao:2001yh} are then used to estimate
the masses of other bottomonium hybrids, which are  
given in Table~\ref{tab:bhybrids}.
The error bars in Table~\ref{tab:bhybrids} take into account 
the statistical errors in the lattice calculations
and the uncertainties from setting the heavy-quark mass.
They do not account for systematic errors associated with extrapolations 
to zero lattice spacing and from the omission of light-quark loops.
The error bars in the mass splittings from the lattice QCD calculation
are much larger than those from the lattice NRQCD calculation. 
They are comparable to the splitting between the $H_1$ and $H_3$ multiplets.

We can compare the mass splittings for bottomonium hybrids
in Table~\ref{tab:bhybrids} calculated using quenched lattice NRQCD
with the mass splittings for charmonium hybrids 
in Table~\ref{tab:chybrids} calculated using lattice QCD with dynamical light quarks.  
The central value of the splitting between the $1^{++}$ state of $H_2$ 
and the $1^{--}$ state of $H_1$ for bottomonium hybrids is 
about 1/3 that for charmonium hybrids.
The central value of the splitting between the $0^{++}$ state of $H_3$ 
and the $1^{--}$ state of $H_1$ for bottomonium hybrids is about twice 
as large as that for charmonium hybrids.
Definitive calculations of the spectrum of bottomonium hybrids 
using lattice NRQCD with dynamical light quarks would be valuable.

\subsection{Charmonium tetraquarks from lattice QCD}
\label{sec:latticetetra}

Lattice QCD has not yet provided much information on quarkonium tetraquarks.
Pelovsek and Leksovec have made a first attempt to observe the 
charmonium tetraquark $Z_c(3900)$ using lattice QCD with dynamical $u$ and $d$ quarks
under the assumption that its $I^G(J^{P})$ quantum numbers 
are $1^+(1^+)$ \cite{Prelovsek:2013xba}.
They looked for a signal for the $Z_c$ in the cross correlators of 6 operators.
Three of the operators were linear combinations of products 
of color-singlet $c \bar q$ and $q \bar c$ operators,
so they couple most strongly to states that consist of a pair of charm mesons
$D^* \bar D$.
The other 3 operators were linear combinations of products 
of color-singlet $c \bar c$ and $q \bar q$ operators,
so they couple most strongly to states that consist of $J/\psi\, \pi$.
The only signals they observed were for scattering states of the meson pairs
$D^* \bar D$ and $J/\psi\, \pi$.

In the Born-Oppenheimer picture, the component of  a charmonium tetraquark 
in which the $c \bar c$ pair is close together can be approximated by a
charmonium adjoint meson, which consists of a light $q \bar q$ pair 
bound to a color-octet $c \bar c$ pair.
This suggests that the operators that couple most strongly to 
charmonium tetraquarks could be linear combinations of products 
of a color-octet $c \bar c$ operator and a color-octet $q \bar q$ operator.
Such operators would have suppressed couplings to scattering states consisting of 
a pair of mesons.
Operators with this structure should be included in a comprehensive study of 
charmonium tetraquarks in lattice QCD.

\subsection{Quarkonium hybrids from QCD sum rules}
\label{sec:sumrules}

Numerous papers have been written in which QCD sum rules are used to postdict
the masses of individual $XYZ$ mesons in Tables~\ref{tab:ccneutral}, \ref{tab:cccharged},
and \ref{tab:bbboth}.  A global analysis of the pattern of $XYZ$ states
predicted by QCD sum rules would be more useful.
Such an analysis has been carried out for charmonium hybrids and bottomonium hybrids
by Chen {\it et al.}\  \cite{Chen:2013zia}.
They identified a total of 10 hybrid states, with the same pattern of masses 
for charmonium hybrids and bottomonium hybrids.
The states with the four lowest masses are those in the multiplet $H_1$ in Eq.~(\ref{H1}).
The five states with the next higher masses have quantum numbers that correspond
to 5 of the 8 states in the multiplets $H_2$, $H_3$, and $H_4$ in Eqs.~(\ref{H1234}).
The highest-mass state identified in Ref.~\cite{Chen:2013zia} 
had the exotic quantum numbers $\bm{0}^{--}$.
In the Born-Oppenheimer approach, the lowest $\bm{0}^{--}$ quarkonium hybrid could be 
a spin-triplet state in the first orbital-angular-momentum excitation of the 
$\Sigma_g^-$ hybrid potential. 
It is useful to compare the QCD sum rule predictions for charmonium hybrids 
with those from lattice QCD in Table~\ref{tab:chybrids}.
The lowest-mass state is predicted by QCD sum rules to be the $1^{--}$ state of $H_1$,
while lattice QCD predicts it to be the $0^{-+}$ state of $H_1$.
Lattice QCD predicts that the higher-mass states can be arranged into multiplets 
$H_2$, $H_3$, and $H_4$ with increasing masses.
QCD sum rules do not predict any such ordering of the masses in the three multiplets.
A $\bm{0}^{--}$ charmonium hybrid was not observed in the lattice QCD calculations 
of Ref.~\cite{Liu:2012ze}.

\section{Phenomenological Analysis}
\label{sec:pheno}

In this section, we present predictions for energy levels of 
charmonium hybrids and bottomonium hybrids using the Born-Oppenheimer approximation
in conjunction with inputs from lattice gauge theory.
We also present a speculative illustration of some of the energy levels of 
charmonium tetraquarks.

\subsection{Quarkonium}
\label{sec:pheno-onium}

A crucial parameter in the B-O approximation is the heavy-quark 
mass $m_Q$, which appears in the radial Schroedinger equation in Eq.~(\ref{scheq4}).
If there was a rigorous derivation of the B-O approximation
as the leading term in a systematic approximation to QCD, 
it would be possible to determine the appropriate value of $m_Q$ 
from the parameters of QCD.
In the absence of such a derivation, an alternative is to treat the 
charm quark mass $m_c$ and the bottom quark mass $m_b$ 
as phenomenological parameters.
Since we wish to determine the energy levels of quarkonium hybrids and 
quarkonium tetraquarks by solving the Schroedinger equation,
we choose to determine $m_c$ and $m_b$ by fitting the quarkonium energy levels 
predicted by the  Schroedinger equation in the $\Sigma_g^+$ potential
to the measured energy levels 
of charmonium and bottomonium.

The energy levels $E_{nL}$ for quarkonium  are eigenvalues of the radial 
Schroedinger equation in Eq.~(\ref{scheq4}) with potential $V_{\Sigma_g^+}(r)$
and with $\Lambda=0$ and $J_\Gamma=0$.  The $\Sigma_g^+$ potential has been 
calculated using lattice QCD.  An obvious way to determine $m_Q$ is to
solve the Schroedinger equation for that potential 
and then fit the single parameter $m_Q$ to the observed quarkonium energy levels.
One problem with this procedure is there are strong correlations
between $m_Q$ and the parameters of the potential.
These parameters include the string tension $\sigma$, 
which determines the flavor-singlet B-O potentials
at large $r$ by Eq.~(\ref{VGamma:largeR}).  The value of $\sigma$ from lattice QCD
has a 10\% error, and this limits the accuracy of the determination of $m_Q$.
To deal with this problem,
we will approximate the $\Sigma_g^+$ potential by the Cornell potential 
in Eq.~(\ref{VCornell-R}) and determine the quark masses $m_c$ and $m_b$
as well as the parameters of the potential, including $\sigma$, by fitting the 
observed energy levels of charmonium and bottomonium.
The fitted value of $\sigma$ will then be used in the parametrization of the 
hybrid potentials.  The energy levels of quarkonium hybrids will be obtained 
by solving the Schroedinger equation for those potentials with the fitted values of 
$m_c$ and $m_b$.  

The quarkonium energy levels in the $\Sigma_g^+$
potential are labeled by a radial quantum number 
$n=1,2,3, \dots$ and by an orbital-angular-momentum quantum number 
$L=0,1,2,\dots$ (or $S,P,D,\dots$). For each energy level $nL$, 
there are multiple states with different $J^{PC}$ quantum numbers 
that are related by heavy-quark spin symmetry. 
The spin-symmetry multiplets for the $\Sigma_g^+(1S)$, $\Sigma_g^+(1P)$, 
and $\Sigma_g^+(1D)$ energy levels are given in Table~\ref{tab:multiplets}.
Three complete charmonium multiplets below the $D\bar D$ threshold 
have been observed: $1S$, $1P$, and $2S$. 
Four complete bottomonium multiplets below the $B \bar B$ threshold 
have been observed: $1S$, $1P$, $2S$, and $2P$. 

\begin{table}[t]
\begin{center}
\begin{tabular}{l|cc}
~ & ~charmonium~ & ~bottomonium~   \\
\hline
~~~~~$1S$~ & ~$3067.9 \pm 0.3$~ & ~~$9445.0 \pm 0.7$~\\
~~~~~$1P$~ & ~$3525.3 \pm 0.1$~ & ~~$9899.9 \pm 0.4$~\\
~~~~~$2S$~ & ~$3674.3 \pm 0.3$~ & ~$10017.2 \pm 1.1$~\\
~~~~~$2P$~ &               ~~               & ~$10260.2 \pm 0.5$~\\
\hline
$m_Q$ (GeV)                         & ~1.48~ & ~4.89~ \\
$\sigma$ (GeV$^2$)              & \multicolumn{2}{c}{0.187} \\
$\kappa$                                & \multicolumn{2}{c}{0.489}  \\
$V_0$ (GeV)~ & \multicolumn{2}{c}{-0.242}  \\
\hline
\end{tabular}
\end{center}
\caption{Spin-averaged energy levels (in MeV) 
for charmonium and bottomonium multiplets
and the parameters of the Cornell potential obtained by fitting those energies
with the constraint $m_b - m_c = 3412.2$~MeV.}
\label{tab:onium-fit}
\end{table}

The Schrodinger equation in the $\Sigma_g^+$ potential predicts 
that the spin states in the multiplet for an energy level $nL$ 
are all degenerate. Spin splittings arise from additional terms 
in the Hamiltonian that can be treated as perturbations. 
The energy levels in the $\Sigma_g^+$ potential can be 
interpreted as averages over the multiplet weighted by the number 
of spin states. The spin-averaged mass for the $1S$ energy level 
of charmonium, $\{ \eta_c(1S),J/\psi \}$, is 
\begin{equation}
M_{c \bar c (1S)} = 
\left( M_{\eta_c(1S)} + 3 M_{J/\psi} \right)/4.
\end{equation}	
For all the observed $P$-wave multiplets, the mass of the 
spin-singlet $1^{+-}$ state is consistent with the spin-weighted average 
of the masses of the spin-triplet states $0^{++}$, $1^{++}$, and $2^{++}$.  
A more precise value for the spin-weighted average mass for the multiplet 
can therefore be obtained by just using the spin-triplet states.  
Thus the spin-averaged mass for the $1P$ energy level of 
charmonium, 
$\{ h_c(1P),(\chi_{c0}(1P),\chi_{c1}(1P),\chi_{c2}(1P)) \}$, 
can be approximated by 
\begin{eqnarray}
M_{c\bar c(1P)} = 
\left( M_{\chi_{c0}(1P)} + 3 M_{\chi_{c1}(1P)} + 5 M_{\chi_{c2}(1P)} \right)/9.  
\end{eqnarray}	
The spin-averaged masses for the $1S$, $1P$, and $2S$ charmonium multiplets 
and the $1S$, $1P$, $2S$, and $2P$ bottomonium 
multiplets are given in Table~\ref{tab:onium-fit}.

The Schroedinger equation for $Q$ and $\bar Q$ interacting through the 
Cornell potential in Eq.~(\ref{VCornell-R}) 
can be solved to obtain the energy levels $E_{nL}$
as functions of the quark mass $m_Q$ and the parameters $\sigma$,
$\kappa$, and $V_0$.
By dimensional analysis, the energy levels have the form
\begin{equation}
E_{nL}^{(Q)} = 2 m_Q + V_0
+ \left( \sigma^2/m_Q \right)^{1/3} \zeta_{nL}(\kappa (m_Q^2/\sigma)^{1/3}).\label{EnL}
\end{equation}	
where $\zeta_{nL}(x)$ is a dimensionless function of its argument.
The additive constant has been assumed to be the sum of $2 m_Q$ 
and a term $V_0$ that is independent of the heavy quark.
The splittings $E_{n'L'}^{(Q)} - E_{nL}^{(Q)}$ between energy levels
depend only on the combinations of parameters
$\sigma^2/m_Q$ and $\kappa (m_Q^2/\sigma)^{1/3}$.
Thus if we only fit the observed energy splittings,
the heavy quark mass is completely arbitrary.
Any change in $m_Q$ can be compensated for all the energy splittings 
simultaneously by changes in $\sigma$ and $\kappa$.
One can determine $2m_Q+V_0$ by subsequently fitting one of the  
energy levels, such as the ground-state energy $E_{1S}^{(Q)}$.  
However, because $2m_Q+m_Q$ is determined by a single measurement only, 
it is more sensitive to the choice of the fitting observable than
the combinations $\sigma^2/m_Q$ and $\kappa (m_Q^2/\sigma)^{1/3}$.

By fitting observed splittings between charmonium energy levels 
and  between bottomonium energy levels simultaneously, one can determine
the combinations $\sigma^2/m_b$ and $\kappa (m_b^2/\sigma)^{1/3}$
and the ratio $m_b/m_c$ of the quark masses, but the individual quark mass
$m_b$ remains completely arbitrary.  The individual quark masses and $V_0$
can be determined by subsequently fitting one of the charmonium energy levels
and one of the bottomonium energy levels, 
such as the ground-state energies $E_{1S}^{(c)}$ and $E_{1S}^{(b)}$.  
However, because it is determined by two measurements only, 
$m_b$ is more sensitive to the choice of the fitting observables
than $m_b/m_c$, $\sigma^2/m_b$, and $\kappa (m_b^2/\sigma)^{1/3}$.  

One way to decrease the sensitivity to the choice of fitting observables
is to impose additional constraints on  the quark masses.
One such constraint is motivated by heavy quark symmetry,
which implies that the mass of a heavy-light meson has an expansion
in powers of $1/m_Q$.    The leading term in the expansion is $m_Q$.
The next-to-leading term of order $m_Q^0$ can be interpreted as the 
constituent mass of the light quark.
This constituent quark mass cancels in the difference between the 
masses of a bottom meson and a charm meson with the same light flavor,
leaving the difference between the quark masses. We choose to determine
the quark mass difference from the difference between the average
of the $B^+$ and $B^0$ masses and the average
of the $D^0$ and $D^+$ masses:
\begin{equation}
m_b - m_c  = m_B - m_D = 3412.2 \pm 0.2~{\rm MeV}.
\label{mbmc}
\end{equation}	
%

\begin{table}[t]
\begin{center}
\begin{tabular}{ccrr|ccrr}
\multicolumn{4}{c}{charmonium} & \multicolumn{4}{c}{bottomonium} \\
~$n\, {}^{2S+1}L_J$~ & ~predicted~ & ~observed~  & ~difference~  &
~$n\, {}^{2S+1}L_J$~ & ~predicted~ & ~observed~  & ~difference~   \\
\hline
$1S$         & 3077  & (3068)~~~ &     +9~~~~~ & $1S$         & ~9442 & ~(9445)~~ & $-3$~~~~~ \\
$1P$         & 3503  & (3525)~~~ &  $-22$~~~~~ & $1P$         & ~9908 & ~(9900)~~ &    +8~~~~~ \\
$2S$         & 3687  & (3674)~~~ &   +13~~~~~ & $2S$         & 10009 & ~(10017)~~ & $-8$~~~~~ \\
$1\, ^3D_1$ & 3802  & 3773~~~~ &   +29~~~~~ & $1\, {}^3D_2$ & 10155 & 10164~~~ &   $-9$~~~~~ \\
$2\, ^3P_2$ & 3976  & 3927~~~~ &   +49~~~~~ & $2P$         & 10265 & ~(10260)~~ &    +5~~~~~ \\
$3\, ^3S_1$ & 4138  & 4039~~~~ &   +99~~~~~ & $3\, {}^3S_1$ & 10356  & 10355~~~ &   +1~~~~~ \\
$2\, ^3D_1$ & 4218 & 4153~~~~ &    +65~~~~~ &                  &             &            &                  \\
$4\, ^3S_1$ & 4525 & 4421~~~~ & +104~~~~~ & $4\, {}^3S_1$ & 10638  & 10579~~~ & +59~~~~~ \\
           &       &      &           & $5\, {}^3S_1$ & 10885  & 10876~~~ & +9~~~~~ \\
           &       &      &           & $6\, {}^3S_1$ & 11110  & 11019~~~ & +91~~~~~ \\
\hline
\end{tabular}
\end{center}
\caption{Energy levels (in MeV) for charmonium and bottomonium.
The energy levels predicted by the Cornell potential with the parameters 
given in Table~\ref{tab:onium-fit} are compared to the central values 
of the observed energy levels from Ref.~\cite{Beringer:1900zz}.
The observed energy levels labelled $nL$ are spin averaged.
They are enclosed in parentheses, indicating that they were used as inputs 
to determine the parameters of the Cornell potential.
The observed energy levels labelled $n\,{}^{2S+1}L_J$ are for individual spin states.}
\label{tab:onium-predict}
\end{table}

We determine the parameters of the Cornell potential model by minimizing the 
$\chi^2$ for the 7 spin-averaged energy levels for charmonium and bottomonium
given in Table~\ref{tab:onium-fit}, with the 7 energy levels equally weighted.
The constraint in Eq.~(\ref{mbmc}) is imposed on the quark masses,
so there only 4 independent parameters: $\sigma$, $\kappa$, $V_0$, and $m_b$.
The resulting values of the parameters are given in Table~\ref{tab:onium-fit}.
The heavy quark masses are $m_c = 1.48$~GeV and $m_b = 4.89$~GeV.
The constituent mass for the $u$ and $d$ quarks can be defined by the difference
between $m_B$ and $m_b$ or between $m_D$ and $m_c$, 
which are equal according to the constraint in Eq.~(\ref{mbmc}).
Using the fitted values for $m_c$ or $m_b$, 
the constituent mass of the light quarks is 390~MeV.
Solving the Schrodinger equation for the Cornell potential
with the parameters given in Table~\ref{tab:onium-fit}, 
we obtain predictions for all the energy levels of charmonium and bottomonium.
The resulting predictions are compared with the observed energy levels in 
Table~\ref{tab:onium-predict}.  For the spin-averaged 
energy levels used in the fit, the differences between the predicted 
and observed energy levels are at most 22~MeV.  
The highest energy levels in both the charmonium 
and bottomonium spectrum are underpredicted by about 100~MeV.
The fitted value of the string tension in Table~\ref{tab:onium-fit},
$\sigma = 0.187~{\rm GeV}^2$,  is compatible to within errors
with the value $0.20 \pm 0.02 ~{\rm GeV}^2$
determined by fitting lattice QCD calculations of the 
$\Sigma_g^+$ potential for 
two flavors of dynamical light quarks \cite{Bali:2000vr}.
The fitted value $\kappa$= 0.489 in Table~\ref{tab:onium-fit}
is significantly larger than the value 0.368 obtained by fitting 
the lattice QCD calculations.

Heavy-quark spin symmetry implies that
the mass splitting between the ground-state spin-triplet and spin-singlet
mesons enters at order $1/m_Q$. 
This suggests that a more accurate constraint on the quark masses
could be obtained by replacing $m_D$ in Eq.~(\ref{mbmc}) 
by the spin-weighted average of the spin-triplet mesons $D^*$ and  
the spin-singlet mesons $D$ and similarly for $m_B$.
This would constrain the quark mass difference to be  3340.5~MeV.
If this constraint is imposed instead of Eq.~(\ref{mbmc}),
the Cornell potential model gives a slightly better fit to the charmonium spectrum,
and it gives a slightly worse fit to the bottomonium spectrum.
However it gives much less reasonable values for the heavy quark masses:
$m_c = 2.60$~GeV and $m_b = 5.94$~GeV.
Given these fitted values of the heavy quark masses, 
the constituent mass for the light quarks is negative.
In light of this difficulty, we choose to use instead the constraint
on the heavy quark masses in Eq.~(\ref{mbmc}).

\subsection{Hybrids}
\label{sec:pheno-hybrid}

The energy levels of quarkonium  hybrids in the B-O approximation
can be calculated by solving the radial Schroedinger equation in Eq.~(\ref{scheq4})
for the hybrid potentials.
An accurate parametrization for the $\Pi_u$  hybrid potentials 
calculated using quenched lattice QCD in Ref.~\cite{Juge:1999ie}
is given by Eq.~(\ref{VGamma:largeR}) with $n_\Gamma = 1$ for $r > r_*$
and by Eq.~(\ref{VGamma:smallR}) for $r < r_*$, 
along with the matching conditions at $r = r_*$ in Eqs.~(\ref{VPimatch}).
An accurate parametrization of the difference between the 
$\Sigma_u^-$ and $\Pi_u$  hybrid potentials for $r <  2.4$~fm is given by
Eq.~(\ref{VSigmau:intR}).  The parameter $r_0$ that appears in 
Eqs.~(\ref{VGamma:smallR}) and (\ref{VSigmau:intR}) 
and in the matching conditions in Eqs.~(\ref{VPimatch})
is the Sommer radius defined by Eq.~(\ref{Sommer}), which
is determined from bottomonium spectroscopy to be $r_0 = 0.50$~fm.
The string tension $\sigma$ enters both potentials through the parametrization of the 
$\Pi_u$ potential for $r > r_*$ in Eq.~(\ref{VGamma:largeR})
and through the matching conditions at $r = r_*$ in Eqs.~(\ref{VPimatch}).
We use the value $\sigma = 0.187~{\rm GeV}^2$ in 
Table~\ref{tab:onium-fit}, which was obtained by fitting the charmonium  
and bottomonium spectra.
The matching point is then determined by Eq.~(\ref{VPimatch2})
to be $r_* = 1.5~r_0$. The difference between the energy offsets
is determined by Eq.~(\ref{VPimatch1}) to be  
$E_{\Pi_u} - E_0 = 2.6~r_0^{-1}$.

\begin{figure}[t]
\centerline{ \includegraphics*[width=16cm,clip=true]{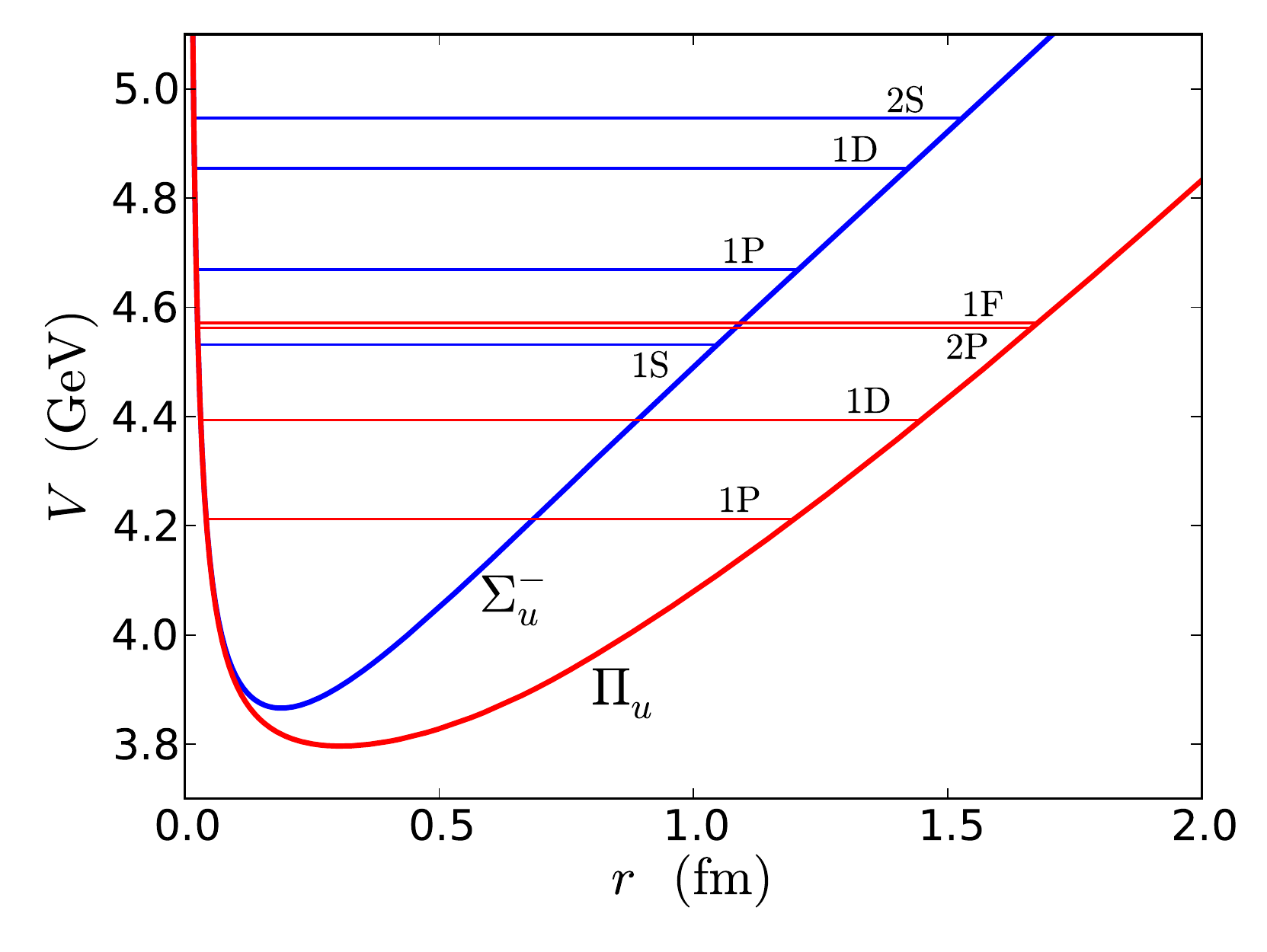} }
\vspace*{0.0cm}
\caption{Lowest energy levels for charmonium  hybrids in the $\Pi_u$ and $\Sigma_u^-$ potentials.
The charm quark mass is $m_c = 1.48$~GeV.
The energy offset has been chosen so that the ground-state $\Pi_u(1P)$ energy level 
is 4.212~GeV.}
\label{fig:chybrids}
\end{figure}

\begin{figure}[t]
\centerline{ \includegraphics*[width=16cm,clip=true]{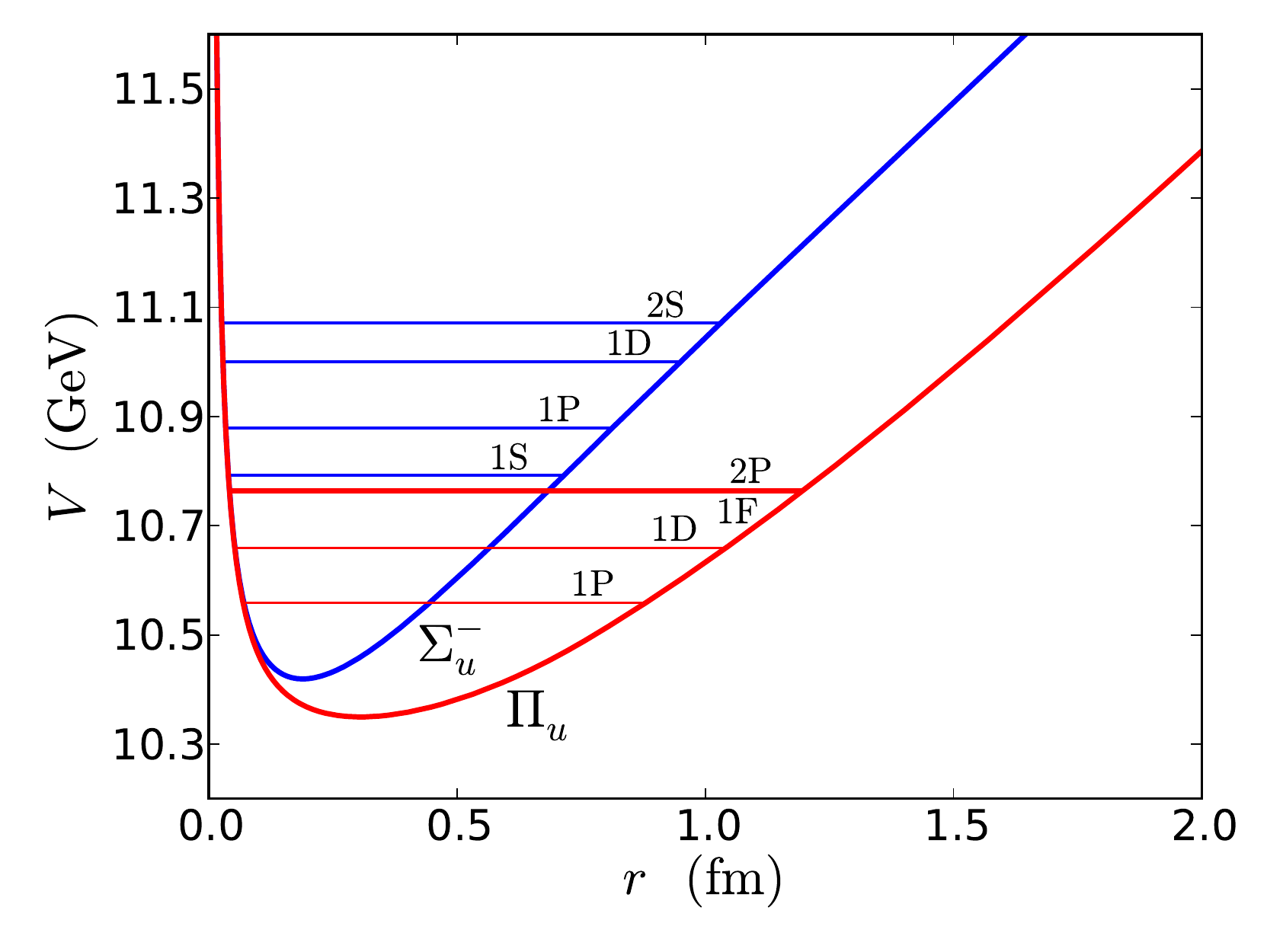} }
\vspace*{0.0cm}
\caption{Lowest energy levels for bottomonium  hybrids in the $\Pi_u$ and $\Sigma_u^-$ potentials.
The bottom quark mass is $m_b = 4.89$~GeV.
The energy offset has been chosen so that the ground-state $\Pi_u(1P)$ energy level
is  10.559~GeV. 
}
\label{fig:bhybrids}
\end{figure}

Given our parametrizations of the $\Pi_u$ and $\Sigma_u^-$ potentials
and given the quark masses $m_c = 1.48$~MeV and $m_b = 4.89$~MeV listed
in Table~\ref{tab:onium-fit}, the quarkonium hybrid energy levels can be obtained by solving the 
radial Schroedinger equation in Eq.~(\ref{scheq4}).  
For charmonium hybrids, 
the ground-state $\Pi_u(1P)$ energy level is predicted to be 4246~MeV, 
and the $\Sigma_u^-(1S)$ energy level is predicted to be higher by 320~MeV.
For bottomonium hybrids, 
the ground-state $\Pi_u(1P)$ energy level is predicted to be  10864~MeV, 
and the  $\Sigma_u^-(1S)$ energy level is predicted to be higher by 233~MeV. 

It is useful to compare the energy levels of charmonium hybrids in the 
B-O approximations with the spin-averaged energy levels 
from lattice QCD in Table~\ref{tab:chybrids}.
In the B-O approximation, the $\Pi_u^+$ and
$\Pi_u^-$ configurations have the same potential $V_{\Pi_u}(r)$,
so they have the same energy levels.
However the lattice QCD result in Table~\ref {tab:chybrids} for
the spin-averaged energy of the $H_2 = \Pi_u^-(1P)$ multiplet 
is about 100~MeV larger than that for the $H_1 =\Pi_u^+(1P)$ multiplet.
With our parametrizations of the $\Pi_u$ and $\Sigma_u^-$ potentials,
the  $\Sigma_u^-(1S)$ energy level for charmonium hybrids 
is predicted to be higher than the  $\Pi_u(1P)$ energy level by 320~MeV. 
However the lattice QCD result in Ref.~\cite{Liu:2012ze} for
the spin-averaged energy of the $H_3 = \Sigma_u^-(1S)$ multiplet
is about 195~MeV higher than that for the $H_1 = \Pi_u^-(1P)$ multiplet.

The spectra of charmonium hybrids and bottomonium hybrids are illustrated 
in Figs.~\ref{fig:chybrids} and \ref{fig:bhybrids}, respectively.
Only the energy levels for the ground state, the first two orbital-angular-momentum 
excitations, and the first radial excitation in each potential are shown.
To facilitate the comparison with later results, we have adjusted the energy offsets 
for the charmonium hybrids and bottomonium hybrids separately.
For charmonium hybrids, we choose the offset so the $\Pi_u(1P)$ energy level 
is equal to 4212~MeV, which is the central value of the spin-averaged energy 
for the $H_1$ multiplet from lattice QCD given in Table~\ref {tab:chybrids}.
For bottomonium hybrids, we arbitrarily choose the offset so the $\Pi_u(1P)$ energy level 
is equal to the $B \bar B$ threshold, which is 10.559~GeV.
The pattern of the energy levels for charmonium hybrids and bottomonium hybrids
in Figs.~\ref{fig:chybrids} and \ref{fig:bhybrids} are quite similar.
The orbital-angular-momentum splittings are smaller than the radial splittings.
The energy splittings relative to the $\Pi_u(1P)$ energy level 
are smaller for bottomonium hybrids  than for charmonium hybrids.
For the energy levels shown in Figs.~\ref{fig:chybrids} and \ref{fig:bhybrids},
the splittings for bottomonium hybrids are less than 3/4 
of those for the corresponding charmonium hybrids.
The bottomonium hybrids are smaller than the corresponding charmonium hybrids.
One measure of the size is the outer classical turning radius at which the 
potential is equal to the energy of the state.
For the energy levels shown in Figs.~\ref{fig:chybrids} and \ref{fig:bhybrids},
the outer classical turning radii for the bottomonium hybrids are less than 3/4 
of those for the corresponding charmonium hybrids.

\begin{table}[t]
\begin{center}
\begin{tabular}{ccc|cccc}
~multiplet~ & $\Gamma$ & ~$nL$~ & ~$c \bar c$ hybrid~ & ~$b \bar b$ hybrid~ & $S=0$ & $S=1$ \\
\hline
$H_1$ & ~$\Pi_u^+$~ & ~$1P$~ & ~(4212)~ & ~(10559)~ & 
~$1^{--}$~ & ~$(0,\bm{1},2)^{-+}$~ \\
$H_4$ &                      & ~$1D$~ &    4394    &    10659   & ~$2^{++}$~ & ~$(1,\bm{2},3)^{+-}$~ \\
$H_1'$ &                     & ~$2P$~ &    4562     &   10766   & ~$1^{--}$~ & ~$(0,\bm{1},2)^{-+}$~ \\
\hline
$H_2$ & ~$\Pi_u^-$~        & ~$1P$~ & ~(4314)~ & ~(10597)~ & 
~$1^{++}$~ & ~$(\bm{0},1,\bm{2})^{+-}$~ \\
            &              & ~$1D$~ &          4496    &     10697   & ~$2^{--}$~ & ~$(\bm{1},2,\bm{3})^{-+}$~ \\
            &             & ~$2P$~ &           4664    &     10804   & ~$1^{++}$~ & ~$(\bm{0},1,\bm{2})^{+-}$~\\
\hline
$H_3$ & ~$\Sigma_u^-$~ & ~$1S$~ & ~(4407)~ & ~(10892)~ & 
~$0^{++}$~ & ~$1^{+-}$~ \\
            &             & ~$1P$~ &           4544     &   10979   & ~$1^{--}$~ & ~$(0,\bm{1},2)^{-+}$~ \\
            &             & ~$2S$~ &           4822     &   11171   & ~$0^{++}$~ & ~$1^{+-}$~ \\
\hline
\end{tabular}
\end{center}
\caption{Energy levels (in MeV) for charmonium and bottomonium hybrids
in the $\Pi_u$ and $\Sigma_u^-$ potentials.
For each configuration $\Gamma$,
the ground-state energy level in parentheses is an input.
The inputs for the charmonium hybrids are the spin-averaged energies 
of the $H_1$, $H_2$, and $H_3$ multiplets
from the lattice QCD results in Table~\ref{tab:chybrids}.
The inputs for the bottomonium hybrids
are the energies of the $1^{--}$, $1^{++}$, and $0^{++}$ states 
from the lattice NRQCD results in Table~\ref{tab:bhybrids}.
The energy splittings for the first orbital-angular-momentum excitation
and the first radial excitation are calculated by solving the radial Schroedinger equation
in the $\Pi_u$ and $\Sigma_u^-$ potentials.  
A boldfaced $\bm{J}$ indicates an exotic quantum number.}
\label{tab:cbhybrids}
\end{table}

We can probably get better estimates for the energies of 
charmonium hybrids by using lattice QCD to determine the ground-state energy for 
each B-O configuration and using the Schroedinger equation only to calculate
the differences between the energy levels $nL$ for that B-O configuration.
As the inputs to determine the ground-state energy levels 
$\Pi_u^+(1P)$, $\Pi_u^-(1P)$, and $\Sigma_u^-(1S)$ for charmonium hybrids, 
we choose the lattice QCD results for the spin-averaged energies 
of the $H_1$, $H_2$, and $H_3$ multiplets in Table~\ref{tab:chybrids}.
As the inputs to determine the ground-state energy levels 
$\Pi_u^+(1P)$, $\Pi_u^-(1P)$, and $\Sigma_u^-(1S)$ for bottomonium hybrids, 
we choose the lattice NRQCD results for the energies of the 
$1^{--}$, $1^{++}$, and $0^{++}$ states in Table~\ref{tab:bhybrids}.
The resulting predictions for the energy levels of 
charmonium hybrids and bottomonium hybrids are given in Table~\ref{tab:cbhybrids}.

\subsection{Tetraquarks}
\label{sec:pheno-tetra}

If the tetraquark potentials were known, we could calculate the energy levels
of quarkonium tetraquarks in the B-O approximation
by solving the radial Schroedinger equation in Eq.~(\ref{scheq4}). 
Unfortunately, our only information about the tetraquark potentials from QCD
is that the lowest-energy adjoint mesons in quenched lattice QCD
are a vector $1^-$ and a pseudoscalar $0^-$.
In Section~\ref{sec:tetrapot},
we inferred from the existence of these adjoint mesons that the deepest tetraquark 
B-O potentials are $\Pi_g$ and $\Sigma_g^+$, which are equal at $r=0$, and $\Sigma_u^-$.
We have no information from QCD about the behavior of these potentials at nonzero $r$.
In order to illustrate the B-O approximation for quarkonium tetraquarks,
we will make the simple assumption that the tetraquark $\Pi_g$ and $\Sigma_g^+$
potentials have the same shapes as the hybrid $\Pi_u$ and $\Sigma_u^-$ potentials,
which are shown in Fig.~\ref{fig:chybrids}.  
Under this assumption, the splittings between energy levels in the tetraquark $\Pi_g$ potential
are the same as those in the hybrid $\Pi_u$ potential,
and the splittings between energy levels in the tetraquark $\Sigma_g^+$ potential
are the same as those in the hybrid $\Sigma_u^-$ potential.
The splittings in the hybrid potentials are given in Table~\ref{tab:cbhybrids}.
If the ground-state energy level for a tetraquark B-O configuration was known,
then all the higher energy levels would be determined.

\begin{table}[t]
\begin{center}
\begin{tabular}{cc|ccccc}
~$\Gamma$~ & ~$nL$~ & ~isospin-1~ & 
~isospin-0~ & ~$s \bar s$~ & ~$S=0$~ & ~$S=1$~ \\
\hline
~$\Pi_g^-$~        & ~$1P$~ & ~{\bf (4023)}~ &  ~{\bf (3918)}~ & ~{\bf (4145)}~  &
~$1^{+-}$~ & ~ $(0,1,2)^{++}$~\\
                           & ~$1D$~ &          4205      &           4100      &          4327      & 
~$2^{-+}$~ & ~$(1,2,3)^{++}$~ \\
                           & ~$2P$~ &          4373      &           4268      &          4495      & 
~$1^{+-}$~ & ~$(0,1,2)^{++}$~\\
\hline
~$\Pi_g^+$~        & ~$1P$~ & {\bf (3898)} & (3793) & (4020) & 
~$\bm{1}^{-+}$~ & ~$(\bm{0},1,2)^{--}$~\\
                           & ~$1D$~ &         4080    &  3975   &  4201  & 
~$\bm{2}^{+-}$~ & ~$(1,2,3)^{++}$~ \\
                           & ~$2P$~ &         4248    &  4143   &  4370  & 
~$\bm{1}^{-+}$~ & ~$(\bm{0},1,2)^{--}$~\\
\hline
~$\Sigma_g^+$~ & ~$1S$~ & (4368) & {\bf (4263)} & (4490) & 
~$0^{-+}$~ & ~$1^{--}$~ \\
                           & ~$1P$~ &   4505  &         4400   &   4627   & 
~$1^{+-}$~ & ~$(0,1,2)^{++}$~\\
                          & ~$2S$~  &   4783   &       4678     &  4905   & 
 ~$0^{-+}$~ & ~$1^{--}$~  \\
\hline
~$\Sigma_u^-$~ & ~$1S$~ & & & & ~$0^{++}$~ & ~$1^{+-}$~ \\
                           & ~$1P$~ & & & & ~$1^{--}$~ & ~$(0,\bm{1}, 2)^{-+}$~\\
                          & ~$2S$~  & & & & ~$0^{++}$~ & ~$1^{+-}$~  \\
\hline
\end{tabular}
\end{center}
\caption{Energy levels (in MeV) for charmonium tetraquarks 
in the $\Pi_g$, $\Sigma_g^+$, and $\Sigma_u^-$ potentials.
The boldfaced experimental inputs in parentheses are the measured masses of the 
$Z_c^+(4020)$, $X(3915)$, $Y(4140)$, $Z_c^+(3900)$, and $Y(4260)$.
The other inputs in parentheses are obtained by assuming that the splittings 
between ground-state energy levels are the same for isospin-1, isospin-0, and $s \bar s$.
The energy splittings for the first orbital-angular-momentum excitation
and the first radial excitation are calculated by solving the Schroedinger equation 
in the tetraquark $\Pi_g$ and $\Sigma_g^+$ potentials under the assumption 
that they have the same shapes as the hybrid $\Pi_u$ and $\Sigma_u^-$ potentials.
For an isospin-1 tetraquark, the $J^{PC}$'s are those of the neutral member 
of the isospin multiplet.  A boldfaced $\bm{J}$ indicates an exotic quantum number.}
\label{tab:cctetra}
\end{table}

In Section~\ref{sec:selection}, selection rules for hadronic transitions were used to
identify some of the $XYZ$ mesons listed in Tables~\ref{tab:ccneutral} and \ref{tab:cccharged}
with ground-state energy levels of charmonium hybrids and charmonium tetraquarks.
The $Z_c^+(4020)$ and $Z_2^+(4050)$ were identified as spin-singlet 
and spin-triplet energy levels in the isospin-1 $\Pi_g^-(1P)$ multiplet, respectively.
The $X(3915)$ and $Y(4140)$ were identified as spin-triplet states
in the isospin-0 and $s \bar s$ $\Pi_g^-(1P)$ multiplets, respectively.
Two possible identifications were proposed for both the 
$Z_c^+(3900)$ and $Y(4260)$.  The $Z_c^+(3900)$ could be a spin-triplet state
in either the isospin-1 $\Pi_g^+(1P)$ multiplet or the isospin-1 $\Sigma_g^+(1S)$ multiplet.
The $Y(4260)$ could be a spin-triplet $1^{--}$ state
in either the isospin-0 $\Pi_g^+(1P)$ multiplet or the isospin-0 $\Sigma_g^+(1S)$ multiplet.
The hybrid $\Pi_u$ potential is deeper than the hybrid $\Sigma_u^-$ potential.
If the tetraquark $\Pi_g$ potential is similarly deeper than the tetraquark $\Sigma_g^+$ potential, 
the more plausible identifications are 
$Z_c^+(3900)$ as the spin-triplet state in the isospin-1 $\Pi_g^+(1P)$ multiplet 
and  $Y(4260)$ as the spin-triplet $1^{--}$ state in the isospin-0 $\Sigma_g^+(1S)$ multiplet.

A speculative illustration of the spectrum of charmonium tetraquarks is 
given in Table~\ref{tab:cctetra}.
The masses of the $Z_c^+(4020)$, $X(3915)$, and $Y(4140)$
are used as the inputs for the $\Pi_g^-(1P)$ energy levels of
the isospin-1, isospin-0, and $s \bar s$ tetraquarks, respectively.
The mass of the $Z_c^+(3900)$ is used as the input for the $\Pi_g^+(1P)$ energy level
of the isospin-1 tetraquark. 
The inputs for the $\Pi_g^+(1P)$ energy levels
of the isospin-0 and $s \bar s$ tetraquarks are then obtained by assuming
that the differences between the $\Pi_g^+(1P)$ and $\Pi_g^-(1P)$ energies
are the same for isospin 1, isospin 0, and $s \bar s$. 
The mass of the $Y(4260)$ is used as the input for the $\Sigma_g^+(1S)$ energy level 
for the isospin-0 tetraquark.
The inputs for the $\Sigma_g^+(1S)$ energy levels
of the isospin-1 and $s \bar s$ tetraquarks are then obtained by assuming
that the differences between the $\Sigma_g^+(1S)$ and $\Pi_g^-(1P)$ energies
are the same for isospin 1, isospin 0, and $s \bar s$. 
In Table~\ref{tab:cctetra}, the energies of the first orbital-angular-momentum excitation
and the first radial excitation for each B-O configuration
were obtained by assuming that their splittings from the ground state 
are the same as the analogous splittings for charmonium hybrids in Table~\ref{tab:cbhybrids}.
The splittings between the $\Pi_g^-$ energy levels in Table~\ref{tab:cctetra} 
and those between the $\Pi_g^+$ energy levels in Table~\ref{tab:cctetra} 
are the same as those between the $\Pi_u^+$ energy levels or between the 
$\Pi_u^-$ energy levels in Table~\ref{tab:cbhybrids}.
The splittings between the $\Sigma_g^+$ energy levels in Table~\ref{tab:cctetra} 
are the same as those between the $\Sigma_u^-$ energy levels in Table~\ref{tab:cbhybrids}.
No results are given in Table~\ref{tab:cctetra} for the energy levels in the $\Sigma_u^-$ 
potential, because there are no $XYZ$ mesons that are plausible candidates 
for any of the $\Sigma_u^-(1P)$ energy levels.
The energy levels in Table~\ref{tab:cctetra} are based on very naive assumptions 
about the tetraquark potentials, so they should be treated as illustrative only.

\section{Outlook}
\label{sec:outlook}

The Born-Oppenheimer (B-O) approximation provides a starting point for a 
coherent description of all the $XYZ$ mesons that is based firmly on QCD.
The basis for the B-O approximation is that an $XYZ$ meson 
contains a heavy $Q \bar Q$ pair,
and the time scale for the evolution of the gluon and light-quark fields
is small compared to that for the motion of the $Q$ and $\bar Q$.
The B-O approximation was first developed by Juge, Kuti, and Morningstar
for flavor-singlet $Q \bar Q$ mesons \cite{Juge:1999ie}, 
which are quarkonium and quarkonium hybrids.
However, it can also be applied to flavor-nonsinglet $Q \bar Q$ mesons,
which are quarkonium tetraquarks \cite{Braaten:2013boa}.
Most of the simple constituent models for the $XYZ$ mesons that have been
proposed can be interpreted as different regions of the $Q \bar Q$ wavefunction 
in the B-O approximation.

The B-O approximation involves an adiabatic approximation that reduces the 
aspects of the problem that involve gluon and light-quark fields to the simpler problem
of calculating B-O potentials, which are the energy levels of the light fields
in the presence of static $Q$ and $\bar Q$ sources.
The  B-O potentials can be calculated using lattice QCD.
In order to develop quantitative phenomenology of the $XYZ$ mesons
based on the B-O approximation, it is important to have calculations  of all the
most relevant B-O potentials using lattice QCD with dynamical light quarks.
Juge, Kuti, and Morningstar calculated many of the flavor-singlet B-O potentials
using quenched lattice QCD \cite{Juge:1999ie}.
There have been some calculations of the two deepest hybrid potentials
using lattice QCD with dynamical light quarks \cite{Bali:2000vr,Bali:2005fu}.
There have been no calculations of tetraquark 
potentials using lattice QCD, but there have been calculations of the 
energies of static adjoint mesons using quenched lattice QCD \cite{Foster:1998wu}.  
We have inferred the $\Lambda_\eta^\epsilon$ quantum numbers of the deepest
tetraquark potentials from the $J^{PC}$ quantum numbers 
of the lowest-energy static adjoint mesons.  
Calculations of the deepest tetraquark  potentials 
are needed to confirm their $\Lambda_\eta^\epsilon$ quantum numbers 
and to determine their behavior as functions of $r$.

The adiabatic approximation reduces the QCD problem of determining the
spectrum of $Q \bar Q$ mesons to a quantum mechanics problem for a $Q \bar Q$ pair
with infinitely many coupled channels.  The B-O approximation involves a 
further single-channel approximation that reduces the problem to a 
Schroedinger equation for a single radial wavefunction.  This single-channel
approximation may be adequate for many of the $XYZ$ mesons.
However it breaks down if the mass of the meson is too close to a threshold
for a pair of heavy mesons.  In this case, it is necessary to take into account 
the coupling to the meson-pair scattering channel.
Near the meson-pair threshold, there will be an avoided crossing between
a B-O potential that increases linearly at large $r$
and one that approaches a constant equal to the meson-pair threshold.
Lattice QCD calculations of B-O potentials in regions near their avoided
crossings are needed in order to determine the effects 
of the couplings between the channels.

The B-O approximation can be used to describe hadronic transitions between $XYZ$ mesons.
The spin selection rule and the B-O selection rules provide strong constraints on the $XYZ$
mesons that are plausible candidates for specific energy levels in the hybrid 
and tetraquark potentials \cite{Braaten:2014ita}.  
If these potentials are calculated as functions of $r$, 
a much more detailed phenomenology of the hadronic transitions can be developed.
Given an observed hadronic transition between energy levels in two B-O potentials,
the rate for the same hadronic transition between any other pair of energy levels 
can be estimated using overlap integrals of radial wavefunctions 
and the group theory for angular momentum.

To understand the $XYZ$ mesons in detail, it will be necessary to develop
a framework in which corrections to the B-O approximation can be calculated systematically.
There is an effective field theory for the $Q \bar Q$ sector of QCD
called potential NRQCD in which the QCD interactions are reduced to 
interaction potentials between the $Q$ and $\bar Q$ 
and multipole couplings of the $Q \bar Q$ pair to soft gluons \cite{Brambilla:1999xf}.
Unfortunately, this effective field theory seems to be applicable
only to the most deeply bound
quarkonium states and perhaps only to the ground-state bottomonium states
$\Upsilon(1S)$ and $\eta_b(1S)$.
The development of an effective field theory in which the adiabatic approximation
emerges as a first approximation would provide a powerful framework 
for describing the $XYZ$ mesons.

The B-O approximation predicts that the observed $XYZ$ mesons are only the tip of 
an iceberg.  There are many more $XYZ$ mesons waiting to be discovered.
The selection rules for hadronic transitions can provide some guidance
for searches for additional $XYZ$ mesons.
New $XYZ$ mesons could be discovered in existing data from the 
$B$-factory experiments Belle and Babar and from the LHC experiments
ATLAS, CMS, and LHCb.  The BESII collaboration is continuing to
discover additional  $c\bar c$ $XYZ$ mesons at BEPC-II.
Even more $XYZ$ mesons should be discovered at the upcoming higher-luminosity  
$B$ factory SuperBelle and at the upcoming higher-luminosity  runs of the LHC.
Precision measurements of some of the properties of  $XYZ$ mesons
should be possible at both SuperBelle and eventually at the PANDA detector at GSI.
All this additional data will make the elucidation of the nature of the $XYZ$ 
mesons almost inevitable.  It will deliver a definitive verdict on whether the 
Born-Oppenheimer approximation provides
a coherent theoretical framework for understanding the $XYZ$ mesons.

\begin{acknowledgments}
This research was supported in part by the Department of Energy 
under grant DE-FG02-05ER15715 and by the National Science Foundation under grant PHY-1310862.
\end{acknowledgments}


%

\end{document}